\documentclass[preprint]{elsarticle}
\usepackage[margin=2.5cm]{geometry}

\pdfoutput=1

\usepackage[outdir=./]{epstopdf}
\usepackage{amsmath}
\usepackage{subfig} 
\usepackage{hyperref} 
\usepackage{multirow} 
\usepackage[export]{adjustbox} 
\usepackage{booktabs} 
\usepackage{lineno}

\begin{document}

\begin{frontmatter}

\title{Thermo-electrical modelling of the ATLAS ITk Strip Detector}
\author[1]{Graham Beck}      
\author[2]{Kurt Brendlinger} 
\author[2]{Yu-Heng Chen}     
\author[3]{Georg Viehhauser} 

\address[1]{Queen Mary University of London, Mile End Road, London E1 4NS, UK}
\address[2]{Deutsches Elektronen-Synchrotron DESY, Notkestra{\ss}e 85, 22607 Hamburg}
\address[3]{University of Oxford, Keble Rd, Oxford OX1 3RH, UK}

\begin{abstract}

In this paper we discuss the use of linked thermal and electrical network models to predict the behaviour of a complex silicon detector system. We use the silicon strip detector for the ATLAS Phase-II upgrade to demonstrate the application of such a model and its performance. With this example, a thermo-electrical model is used to test design choices, validate specifications, predict key operational parameters such as cooling system requirements, and optimize operational aspects like the temperature profile over the lifetime of the experiment. The model can reveal insights into the interplay of conditions and components in the silicon module, and it is a valuable tool for estimating the headroom to thermal runaway, all with very moderate computational effort.

\end{abstract}


\end{frontmatter}

\section{Introduction}
The temperatures in silicon detector systems are critically important to their performance. Fundamentally, the leakage current of a silicon sensor has a pronounced temperature dependence 
\begin{equation}
I\propto T_\text{S}^2e^{-T_\text{A}/T_\text{S}},
\label{eq:leakage_current_temp_dependence}
\end{equation}
where $T_\text{S}$ is the sensor temperature and $T_\text{A}\simeq7000$~K \cite{Chilingarov_2013}. Leakage currents in the silicon sensor can become particularly significant after irradiation, and the heat generated by these leakage currents, together with the heat from front-end electronic components on the detector, needs to be removed by a cooling system. The capability of the cooling system to remove this heat is limited by the temperature of the local cold sink (typically a circulated fluid) and the thermal impedance of the heat path between the source (electronics and sensor) and the sink. Due to the strong growth of leakage power with temperature, there is a critical temperature $T_\text{crit}$ above which the heat cannot be removed quickly enough, and the detector becomes thermally unstable (`thermal runaway')\footnote{In a real detector system, the resulting growth of sensor temperature would be arrested by overcurrent limits in the power supplies, resulting in a reduction of the bias voltage. At the same time, the increased current leads to an increase of the noise, such that the overall result is a degradation of the S/N performance of the system.}. Understanding the thermal behaviour and the headroom to thermal runaway is crucial for the design of a silicon detector system. Even before the limit of thermal stability is reached, temperatures in silicon detector systems have a major impact on key system parameters such as power supply capacity and cable dimensions, necessitating an accurate estimate.

In addition to the silicon,
there can be aspects of the front-end electronics that have a temperature dependence. In the strip system for the ATLAS Phase-II upgrade \cite{Collaboration:2017mtb}, which is the subject of this case study, there are two additional temperature-dependent heat sources. The first is a radiation damage effect in the readout electronics, which leads to an increase in the digital power of the chip whose magnitude depends on the total ionisation dose (TID) and the temperature of the chip \cite{Collaboration:2017mtb}. This phenomenon was first observed in the ATLAS IBL \cite{ATL-INDET-PUB-2017-001}. The other temperature dependence of a power source stems from the converter chip (FEAST \cite{1748-0221-6-11-C11035,dcdc-info}) used in the on-detector DC-DC converter system supplying power to the front-end electronics. The FEAST chip has an efficiency that decreases at higher temperatures; its efficiency also depends on the magnitude of the load current.

In principle, the temperatures in the system for a given set of operational parameters (power density, thermal conductivities, etc.) can be predicted using finite element analysis (FEA) to an accuracy that is limited only by the quality of the input parameters. However, this is a time-consuming process and can be prohibitively difficult if a number of local heat sources depend non-linearly on temperature. A simplification to this problem that allows for an analytical solution in the case of a simple heat source topology has been developed in~\cite{Beck:2010zzd}. Here we develop this method further to include several temperature-dependent non-linear heat sources in the front-end electronics. The resulting set of equations cannot be solved analytically anymore, but the solution can be found with little effort using numerical problem solvers. This enables us to predict with some confidence the temperatures and power requirements in the ATLAS strip system throughout Phase-II operation. The results from this prediction have been used throughout the ATLAS strip project to consistently dimension the different systems (cooling, power, services, etc.), including an appropriate margin due to the inclusion of a common set of safety factors. This method can be easily adapted to any other system by adjusting the model to the system-specific geometries and parameters.

\subsection{The ATLAS strip system}
The strip system for the ATLAS Phase-II upgrade consists of two parts: the barrel system, comprised of four concentric cylindrical barrels, and two endcaps consisting of six disks each.

The basic unit of the detector is the silicon-strip module. Modules are composed of sensors with hybrids on top, which host the front-end chips as well as circuitry to convert the supply voltage to the chip voltages. The modules are glued onto both sides of a composite sandwich (local support) that contains two parallel thin-wall titanium cooling pipes embedded in carbon foam (Allcomp K9) between two facesheets of ultra-high-modulus carbon fibre (3 layers of K13C2U/EX1515) co-cured together with a Kapton/copper low-mass tape. During operation, cooling will be achieved by evaporating CO$_2$ in the cooling pipes with a final target temperature no higher than $-35~^\circ$C anywhere along the local support.

\begin{figure}[t!]
\centering
\includegraphics[width=0.8\linewidth]{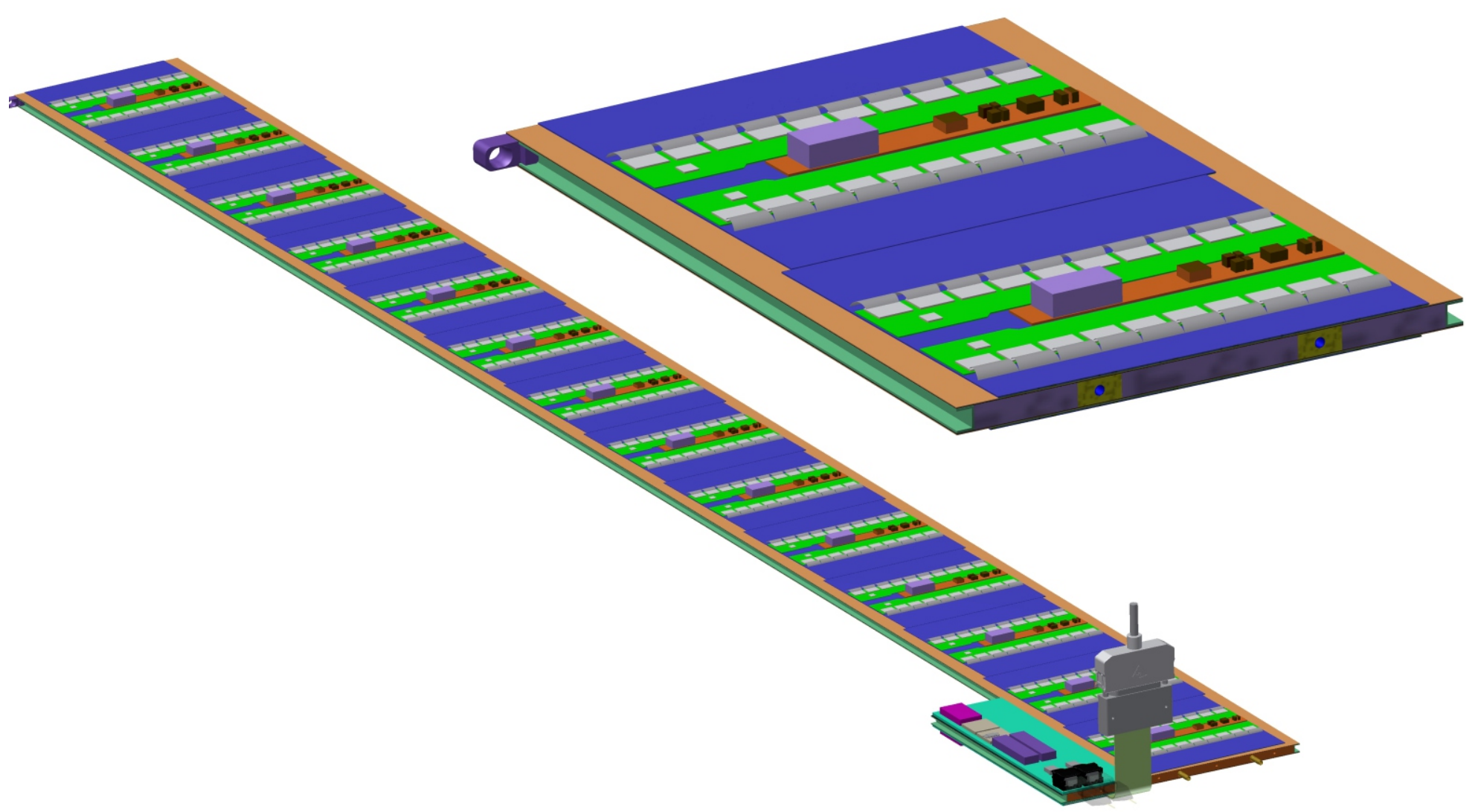}
\caption{Strip barrel local support geometry. On the left, a complete stave is shown (EOS card in the foreground). The right picture shows a cross-section of the stave with the two cooling pipes visible inside the core. }
\label{fig:barrelgeometry}
\end{figure}

In the barrel, the local support is called a stave, onto which 14 modules made with square sensors ($96.85\times 96.72$~mm$^2$) are loaded. Two types of module are used in the barrel: modules with `short-strip' sensors of length 24.1~mm, and `long-strip' sensor modules with 48.2~mm strip lengths. Short-strip modules are used in the two innermost barrel layers, and long-strip modules in the two outermost barrel layers, mainly for hit occupancy considerations. A model of the short-strip barrel stave geometry is shown in Fig.~\ref{fig:barrelgeometry}.

The geometry of the barrel stave is uniform along its length, with the exception of the end region of the stave, where an End-Of-Substructure (EOS) card is mounted on both surfaces. The EOS card shares part of its heat path with the adjacent module; underneath this module (hereafter referred to as an `EOS module'), the thermal path is degraded by the presence of electrically-insulating ceramic pipe sections. The thermal and electrical properties of an EOS module are sufficiently different from other modules along the length of the stave (`normal modules') to warrant separate treatment in the thermo-electrical model of the barrel.

\begin{figure}[t!]
\centering
\includegraphics[width=0.8\linewidth]{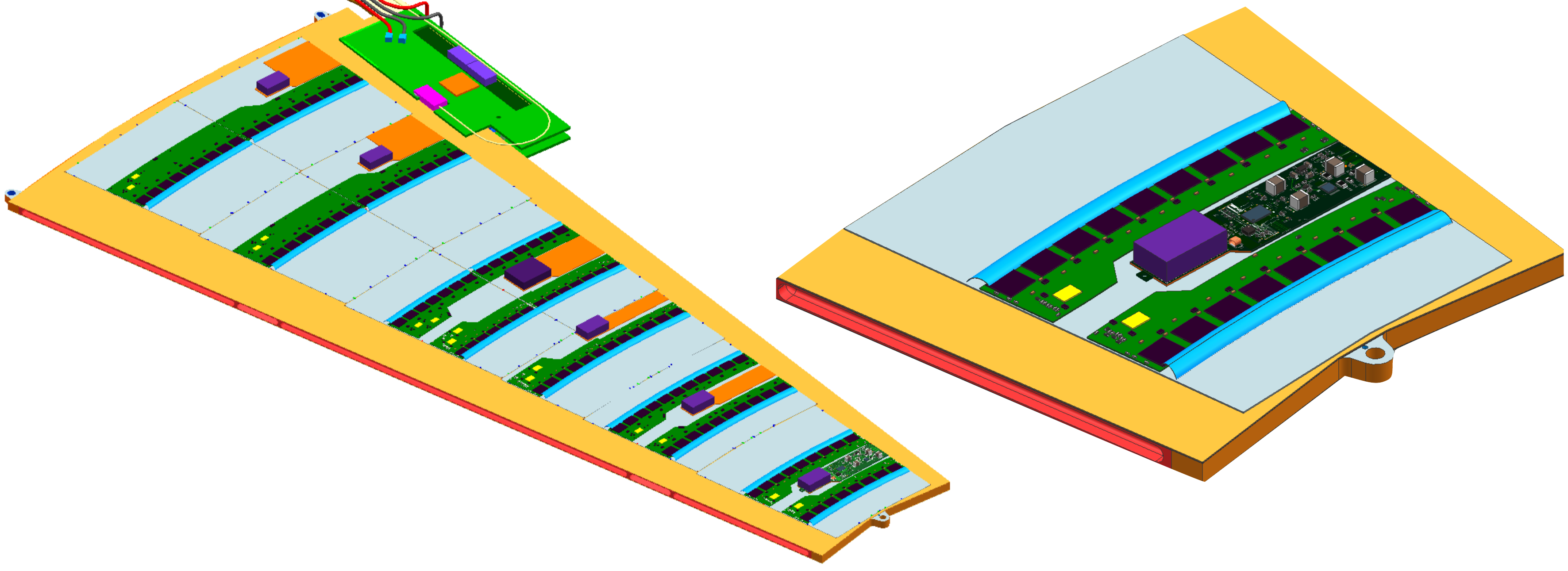}
\caption{The geometry of the endcap strip petal, featuring 6 distinct module designs. A close-up of the R0 module is shown on the right.}
\label{fig:endcapgeometry}
\end{figure}

The endcap system consists of two endcaps composed of 6 disks each.
Each disk contains 32 `petals,' the local substructure depicted in Fig.~\ref{fig:endcapgeometry}.
Both sides of the petal are loaded with 6 modules, each with a distinct design,
located at increasing radius from the beam pipe and labelled R0 through R5 (where `R' stands for ring).
Each endcap module consists of one
or two wedge-shaped silicon sensors and a varying number of front-end chips and DC-DC converters.
The EOS card is located adjacent to the R5 module, but the
cooling pipes run directly underneath it without a shared heat path, in contrast to the barrel EOS card.
Because of the unique geometry of each module in a petal, each of the six different types of module is
modelled separately in the thermo-electrical model.

\subsection{Radiation environment}
A key input to the thermo-electrical calculation is the radiation environment of the strip system, as several inputs depend on radiation damage effects. The sensor leakage current can be parametrized as  a function of the fluence expressed in 1~MeV neutron-equivalents, and the TID effect on the digital chip current will be described as a function of the total ionizing dose rate (more details on its dependencies can be found in Section~\ref{fig:feast_eff}). 

\begin{figure}[ht]
\centering
\subfloat[] {\includegraphics[width=0.48\linewidth]{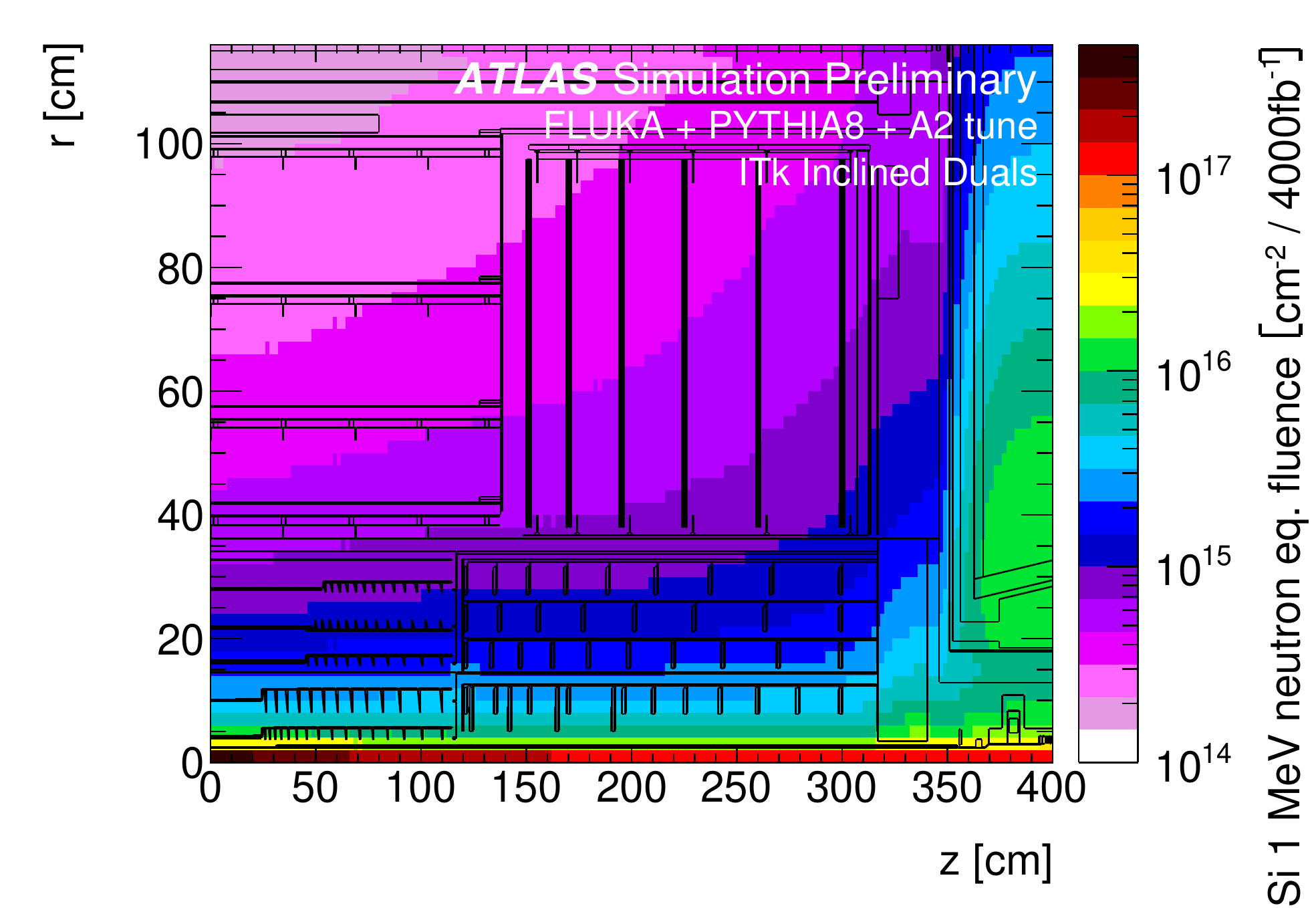}}\quad
\subfloat[] {\includegraphics[width=0.48\linewidth]{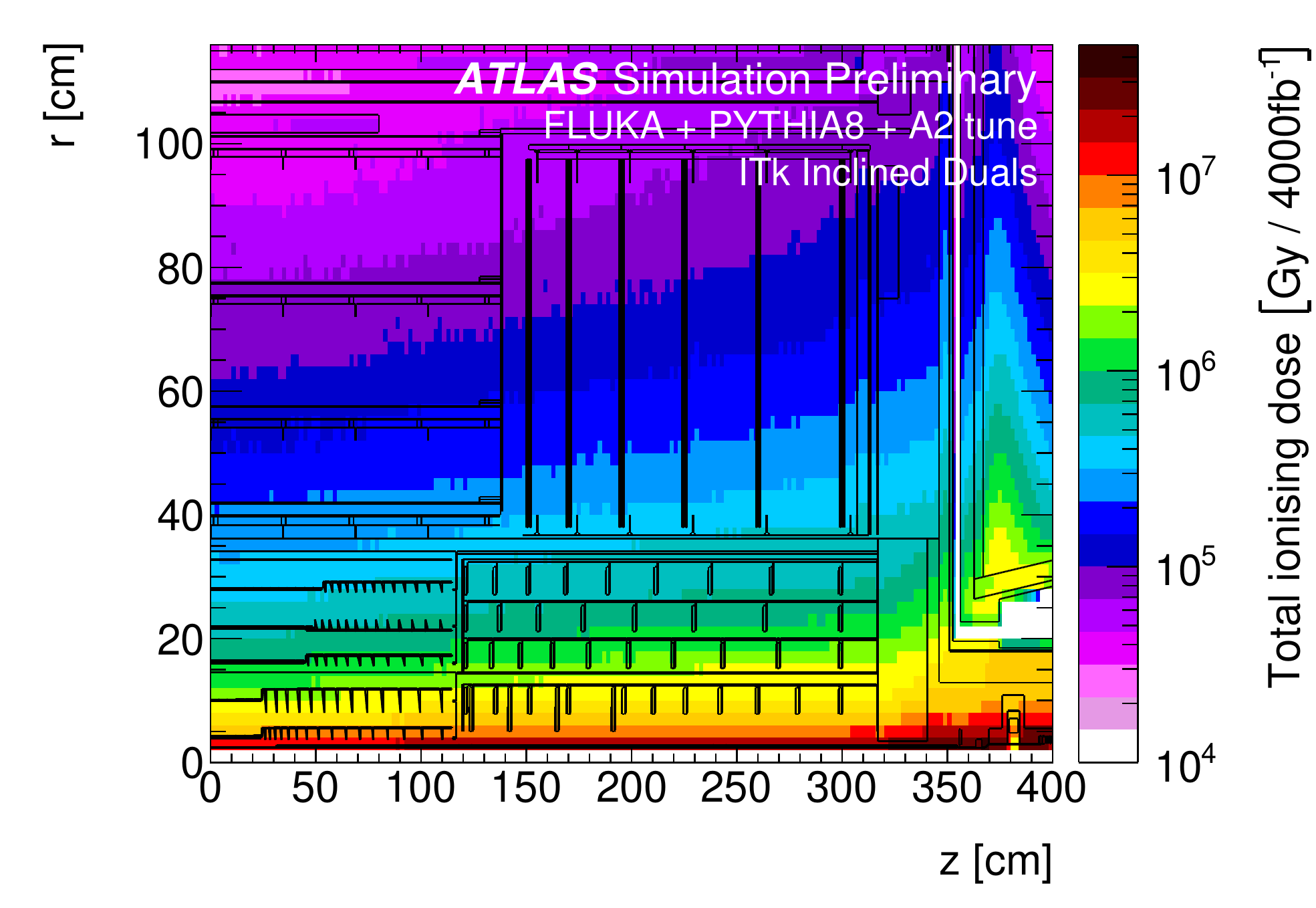}}
\caption{The ATLAS ITk radiation environment. (a) 1~MeV neutron equivalent fluence and (b) total ionizing dose. Both plots are for an integrated luminosity of 4000~fb$^{-1}$ \cite{background}.
The figures use a rainbow colour gradient, with violet indicating the lowest values and red the highest values of fluence or ionizing dose.
}
\label{fig:radiation}
\end{figure}

Predictions for both of these quantities have been generated for each point in the ITk using the FLUKA particle transport code and the PYTHIA8 event generator (Fig.~\ref{fig:radiation}) \cite{background}. In the barrel system, both of these distributions display a strong dependence on $r$ but a weak $z$-dependence. Accordingly, we make the simplifying assumption that modules within the same barrel layer have identical fluence and TID, and model four different radiation profiles (one for each barrel layer). In the endcaps, the radiation levels vary significantly over the length of the petals and from disk to disk; therefore, we model each disk and ring position separately (36 in total).

\section{The electrical model}

\begin{figure}[t!]
\centering
\includegraphics[width=0.8\linewidth]{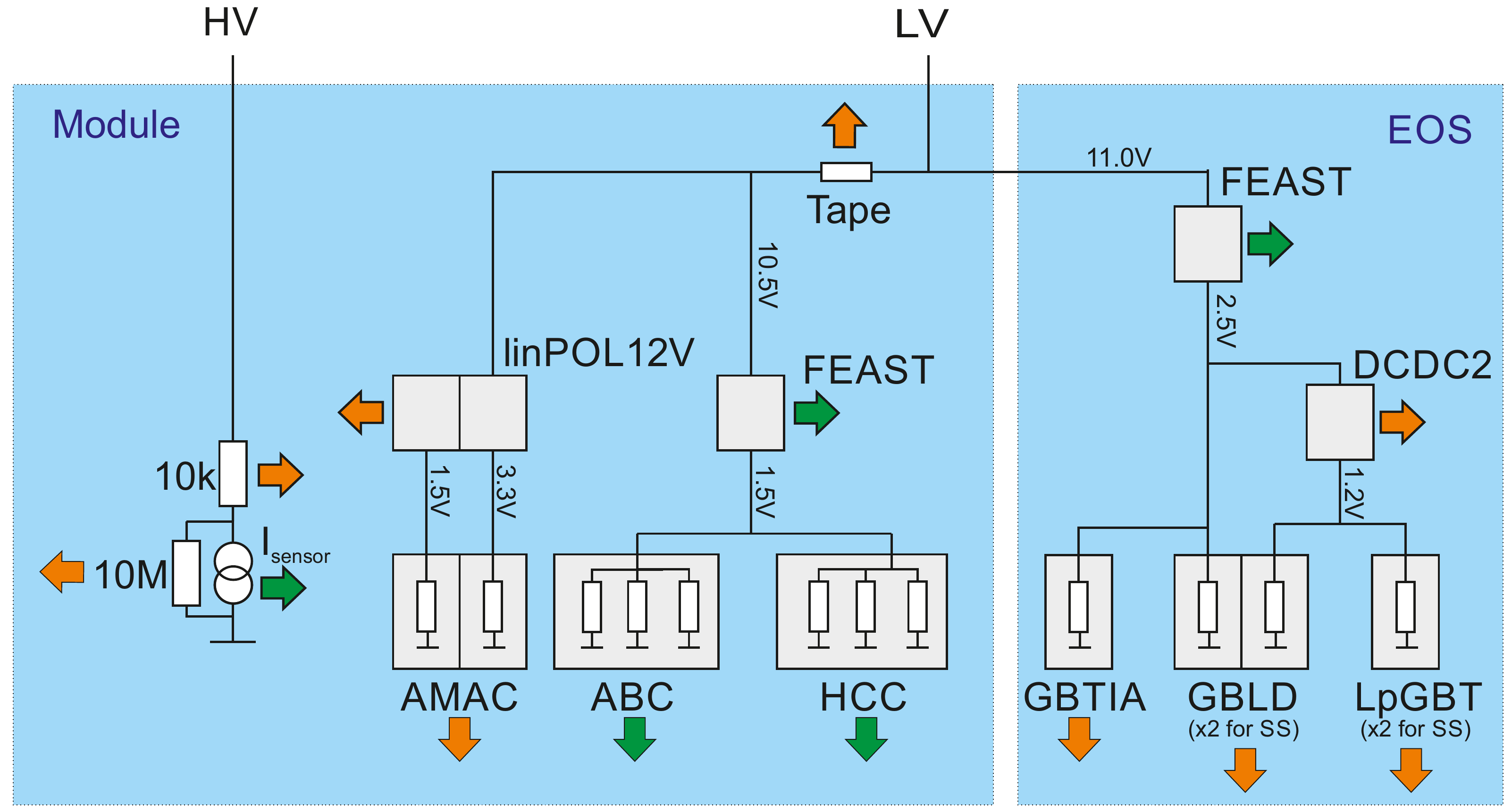}
\caption{
The electrical model of the ITk Strip barrel and endcap modules. Green arrows represent temperature-dependent heat sources, while orange arrows are temperature-independent. Grey squares are chips.
}
\label{electrical_model}
\end{figure}

The electrical model consists of low-voltage (LV) and high-voltage (HV) circuits, depicted in
Fig.~\ref{electrical_model}. The LV current (supplied at 11~V) is used to power the hybrid controller chips (HCCs) \cite{Collaboration:2017mtb},
ATLAS Binary Chips (ABCs) \cite{abc130} and Autonomous Monitoring and Control chip (AMAC) located on PCBs that are
glued directly onto the surface of the sensor.
These chips require between 1.5 and 3.3~V, which are provided by the temperature-dependent
FEAST DC-DC converter\footnote{
In the final version of the ITk strip system the FEAST chip is meant to be replaced by a chip
with similar functionality and behaviour, called the bPOL12V chip; we use the term FEAST
throughout this paper.} and an LDO (low-dropout) regulator (labelled linPOL12V in Fig.~\ref{electrical_model}).
The number of chips and converters on each module vary according to the design of
each different module type (barrel short-strip and long-strip modules, and six different endcap
module designs).
A barrel or endcap module contains 10--28 ABC chips, 1--4 HCCs, and
1--2 of each of the other components (linPOL12V/FEAST/AMAC).

The LV current is also delivered to the EOS card to power various data transfer components:
the Gigabit Laser Driver (GBLD), low power GigaBit Transceiver (LpGBT) and Gigabit Trans-impedance Amplifier (GBTIA) chips.
A FEAST identical to the one used on the module is used to step the
voltage down from 11~V to 2.5~V, and an additional DC-DC converter (`DCDC2') brings the voltage down further for some components.
The short-strip barrel staves contain two GBLD and LpGBT chips.

The bus tape, which carries both LV and HV currents, has a small ohmic resistance, which impacts the
module in two ways. First,
the tape itself will generate some heat according to the amount of current passing through it; this
source of heat is accounted for in the model, however the contribution to the total module power
is negligible.
Second, due to the voltage loss along the traces, there is a slight reduction in voltage supplied to successive modules along the substructure.
The treatment of this effect is slightly different in the barrel and
endcap models: in the barrel, the voltage delivered to every module is averaged to 10.5~V; in the endcap,
the $\Delta V$ is estimated based on the calculated expected power loss along the tape for each module and varies between 10.8 and 11~V.
In both the barrel and endcap systems, the impact of using a different treatment is small.

Finally, the HV current provides the bias voltage on the silicon sensors. An HV multiplexer
switch (HVMUX) can be used to disconnect the sensor from the
bias line (it requires a 10~M$\Omega$ resistor parallel to the sensor in order to function). Two HV filters with an effective resistance of 10~k$\Omega$ are situated in series with the
sensor. The nominal operating voltage of the sensor is expected to be 500V, but the system is designed
to handle a bias voltage of up to 700V.

\section{The thermal model}
The thermal network consists of heat sources (some of which are temperature-dependent) and thermal resistances. The latter are given by the properties of the mechanical design (heat conductivities of the materials) and the geometry of the heat path. The geometry is generally 3-dimensional, but it is the strategy of the simple network models to lump the 3D behaviour into one thermal resistance parameter. In the models discussed here, we have used a granularity corresponding to single detector modules for which the thermal resistance has been modelled. The temperatures in the model are then given for the nodes in the network in analogy to the potentials in an electrical network\footnote{The similarity of the electrical and thermal networks we are using in our approach has been exploited before: historically, Fourier's description of heat conduction pre-dated and inspired Ohm's work on electrical resistive networks. Here we followed the opposite direction.}.

\begin{figure}[t!]
\centering
\includegraphics[width=0.6\linewidth]{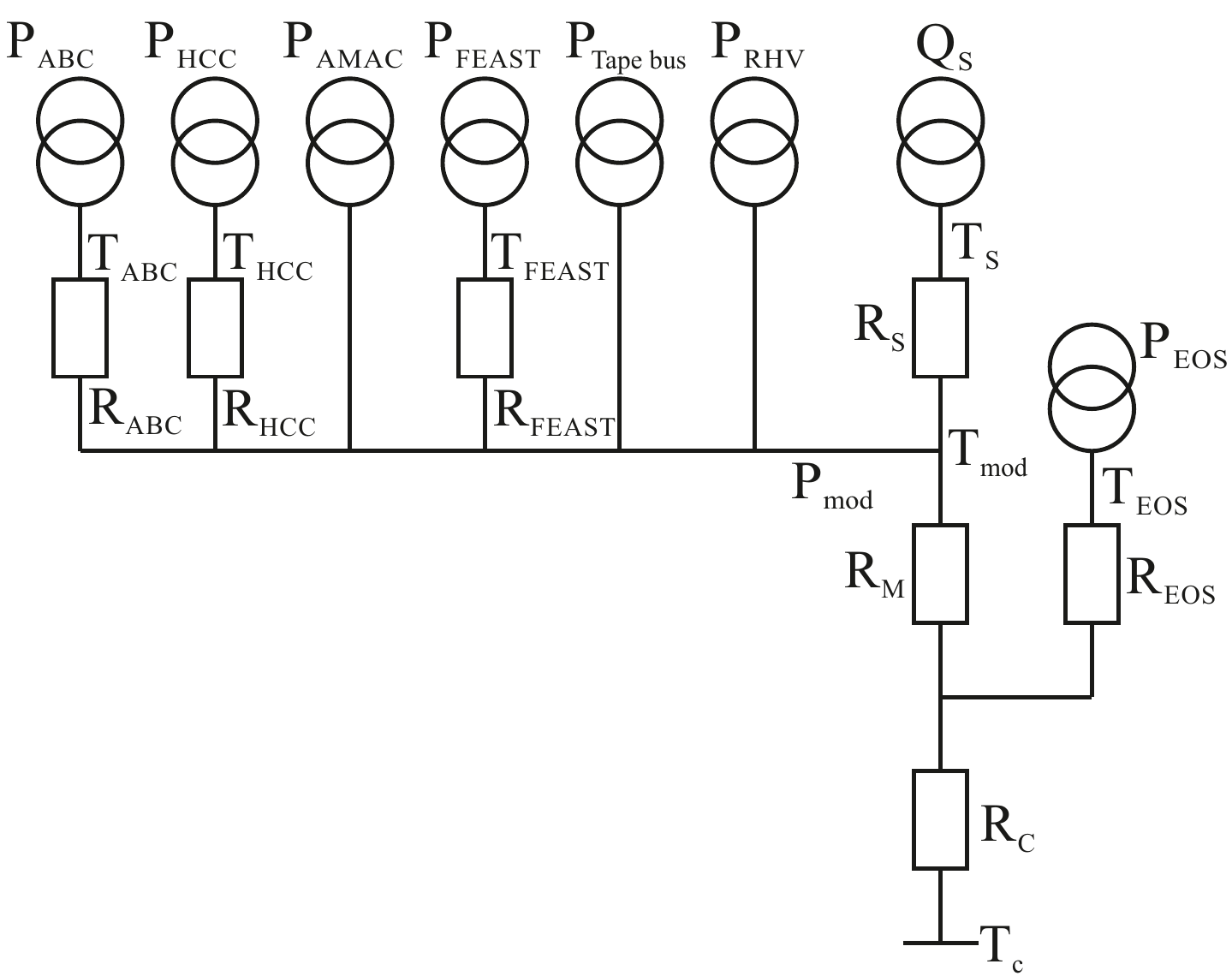}
\caption{Thermal network model.}
\label{fig:thermalmodel}
\end{figure}

The complexity of the thermal network used in this study, depicted in Fig.~\ref{fig:thermalmodel}, is given by the variety of temperature-dependent heat sources in the ATLAS strip system: the digital power for each type of chip, the FEAST chip providing the on-detector DC-DC conversion, and the sensor leakage power. In the ATLAS ITk strip modules, all of these components are located on top of the sensors, such that the heat generated in them flows through the sensor into the support structure, the stave (barrel) or petal (endcap) core with the embedded cooling pipe. In the network model, the heat flow from these sources combines and travels through a common impedance $R_\text{M}$ to the sink at a temperature $T_\text{C}$. For each of the temperature-dependent heat sources (ABC, HCC, FEAST and the sensor) we have added a resistance from the common temperature $T_\text{mod}$ to allow for a finite and different heat path for each of them. Finally, the EOS card adjacent to the last module on the barrel stave or endcap petal is modeled as an additional source of heat with an independent impedance for its unique thermal path.

This is a more complex thermal network than the one studied in Ref.~\cite{Beck:2010zzd}, for which an analytical solution for the determination of thermal stability is given. In particular, because of the non-linear temperature dependence of some of the heat sources, it is not possible in the present case to solve the set of equations describing the model analytically. However, the set of equations is still sufficiently small to solve numerically using any modern mathematical software. We do not expect the numerical method to introduce a loss of precision on the scale of the model approximations.

\section{Obtaining thermal impedances using FEA}
\label{sec:impedances}

The cooling path between the sources dissipating electrical power and the cooling fluid is 3-dimensional and includes components with orthotropic thermal conductivity. Hence the prediction of temperature at any node of the model requires a 3D thermal FEA \cite{abaqus,ansys}. However, the thermal conductivities of the components along the path are approximately constant, so that the temperature rise $\Delta T_i$ above the coolant temperature of any node $i$ ($i=$~ABC, HCC, AMAC, FEAST, tape, RHV, or sensor) in the thermal network model is adequately described by a linear sum of contributions from individual sources, i.e:

\begin{equation}
\Delta T_i \equiv T_i - T_C = R_i P_i + \left(R_\text{C} + R_\text{M}\right)\sum_j P_j,
\label{eq:deltaT_expression}
\end{equation}
where the index $j$ runs over all powered nodes. (We have momentarily disregarded the contribution from the EOS card.)
The expression includes an impedance term $R_\text{CM} \equiv R_\text{C} + R_\text{M}$, the common thermal pathway shared
by the components. We omit the sensor thermal impedance in the definition of $R_\text{CM}$ because its contribution
is negligible (roughly 0.02~$^\circ$C/W).

In order to extract the thermal impedances for the thermal network model, the finite element model is run multiple times, with each heat source (or group of similar sources) switched on in turn with a representative amount of heat. In each of these cases, the temperature is calculated for all nodes in the thermal network model (Fig.~\ref{fig:thermalmodel}). The temperature of a node is here taken as the average of the temperatures for all the grid points in the FEA model within the volume of the object corresponding to the node\footnote{This is particularly interesting in the case of the sensor, which fills a large volume, with a potentially large range of temperatures. In Ref.~\cite{Beck:2010zzd} the analytic model parameters were extracted from the maximum sensor temperature predicted by FEA. Further comparisons after the publication of that paper indicated that the thermal stability limit is predicted more accurately if the average sensor temperature is used.}. The thermal impedances are then obtained from a fit of Eq.~\ref{eq:deltaT_expression} using the temperature data for all nodes for all cases of heat injection.

\begin{figure}[t!]
\centering
\subfloat[Barrel short-strip module] {\includegraphics[width=0.45\linewidth]{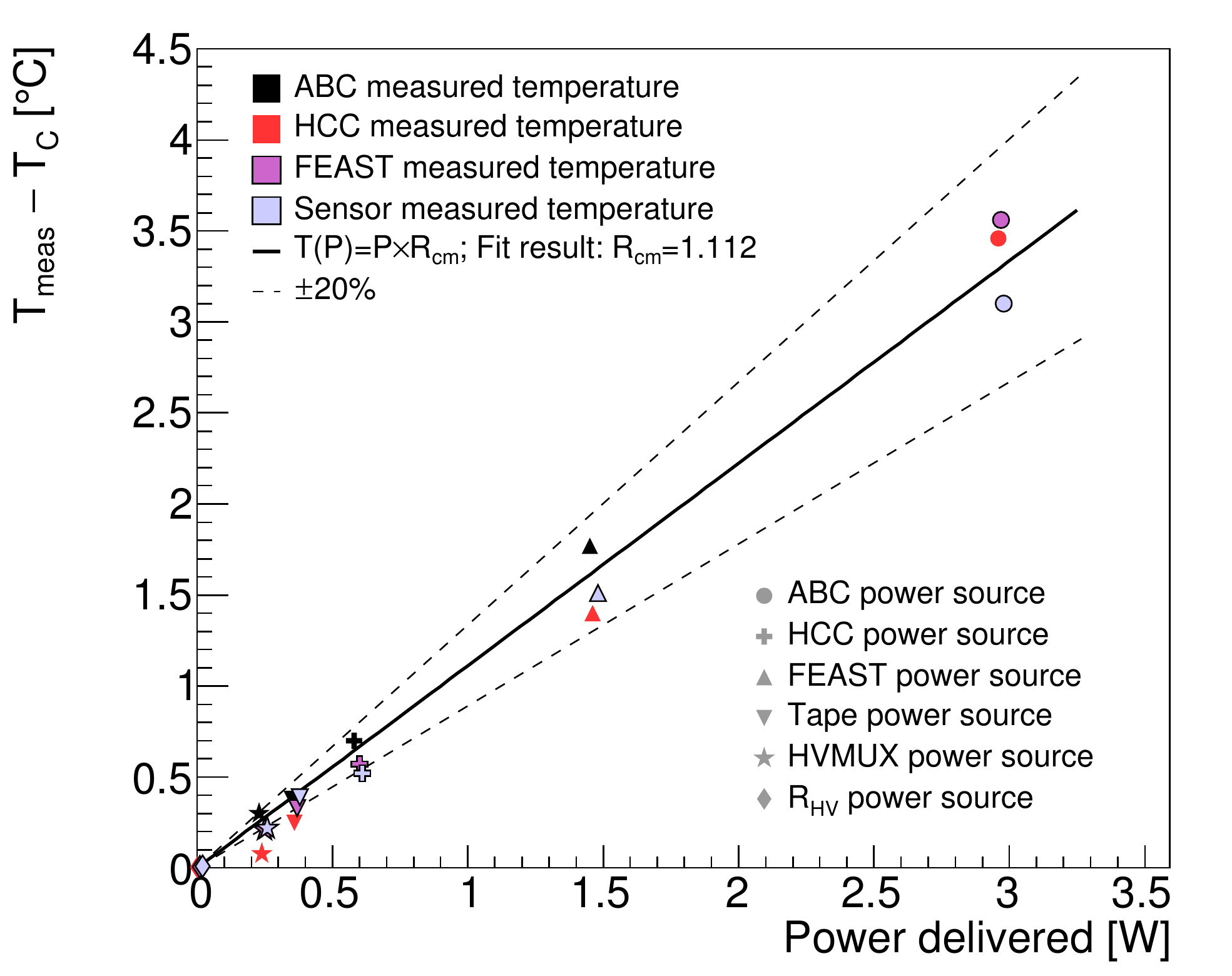}}\quad\quad
\subfloat[Endcap R0 module] {\includegraphics[width=0.45\linewidth]{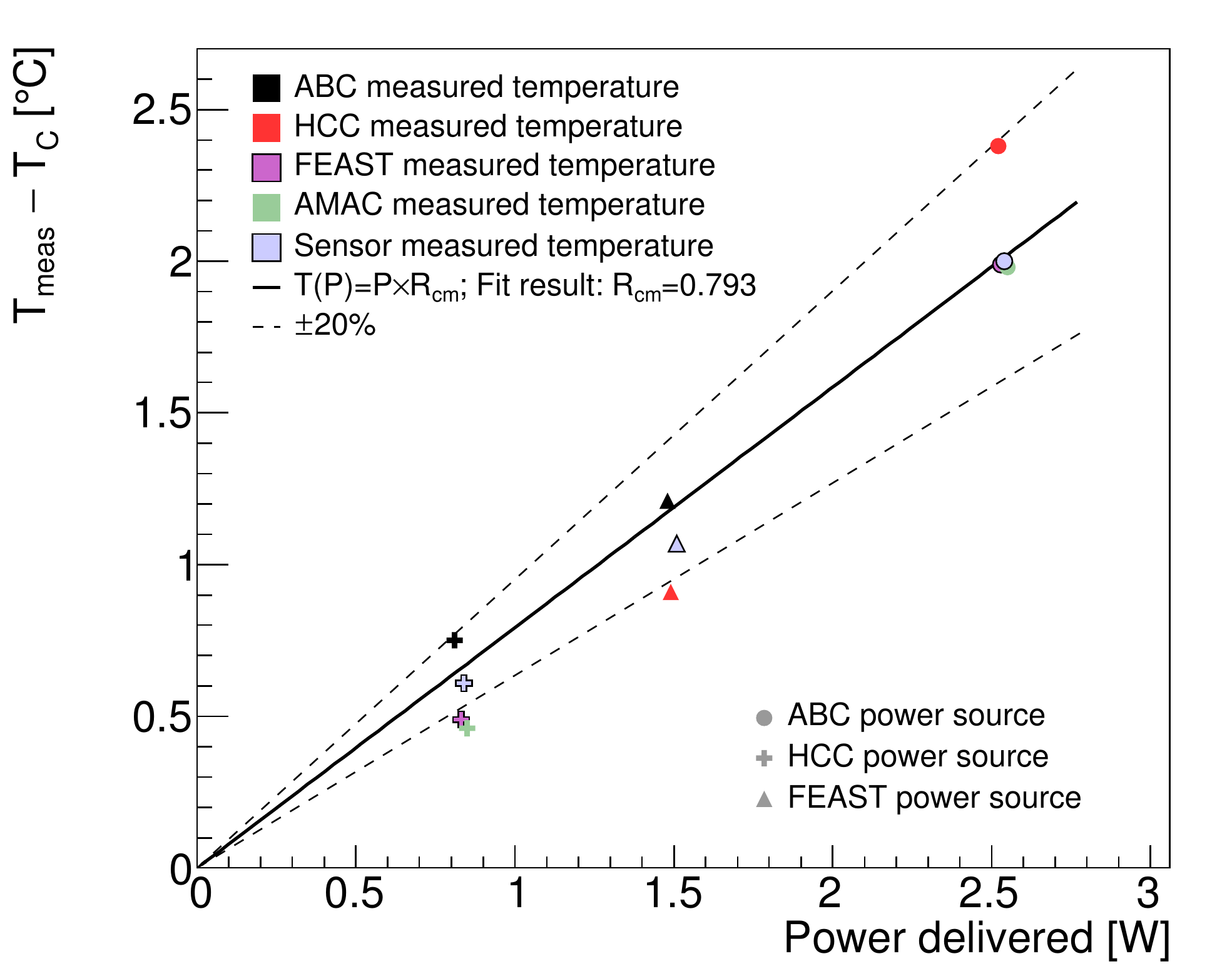}}
\caption{The relationship between the temperature rise observed in the FEA for a specific component and the heat injected in another component. The slope of the fitted line is the estimate for $R_\text{CM}$.
(a) The fit for a short-strip barrel module adjacent to the EOS card. (b) The fit for the endcap R0 module.
For each data point marker, the source of power is indicated by the shape, and the measured component is indicated
by the colour (black: ABC, red: HCC, magenta: FEAST, green: AMAC, light blue: silicon sensor).
The dashed lines represent a $\pm$20\% error band on the fit for $R_\text{CM}$.
}
\label{fig:solving_for_Rcm}
\end{figure}

Because of the nature of the network, the fitted value for the common impedance $R_\text{CM}$ is determined by the observed temperature rises of components where no heat is injected. The linearity of this relationship is illustrated in Fig.~\ref{fig:solving_for_Rcm}. The value of each component-specific impedance is determined from the temperature rise observed when heat is injected into that component. The linear approximation of the model reasonably describes the FEA simulation, and the level of disagreement, discussed below, is taken as an uncertainty that is assessed as a safety factor as described in Section~\ref{sec:safety_factors}.

For the barrel, this procedure is performed for both an EOS module and a normal module, for short-strip and long-strip module types and for 6 different heat sources (24 module FEA simulations in total). In all cases, the agreement of the network temperatures using the thermal impedances from the fit with the data from FEA is better than 0.5~$^\circ$C for all nodes. The thermal impedance from the sensor to the sink ($R_\text{CM}$) is consistently between 1.1 and 1.4~$^\circ$C/W, but higher values (between 10 and 20~$^\circ$C/W) are found for other impedances in the network ($R_\text{HCC}$ and $R_\text{FEAST}$), mostly because these are for components with a small footprint constituting a bottleneck for the heat flow.
A 10\% uncertainty is assigned to the thermal impedances in the barrel modules for the purposes of assessing safety factors.

For the endcap, the procedure to determine the thermal impedances
for each of the 6 module types used simulations of an endcap petal with 3 different heat sources (requiring 3 FEA simulations of a full petal).
$R_\text{CM}$ ranges from 0.6 to 1.4~$^\circ$C/W, with other nodes
between 5 and 20~$^\circ$C/W. Because the location of powered components is more irregular on an
endcap module, the difference between the predicted temperatures of the linear network and the FEA
can reach up to 1.2~$^\circ$C for key temperature-dependent nodes. Therefore, a 20\% uncertainty is assigned
to the thermal impedances in the endcap modules.

There are two recognised departures from linearity of the thermal path: the rise in thermal conductivity of the silicon sensor with decreasing temperature, and the rise in heat transfer coefficient (HTC) of the evaporating CO$_2$ coolant with increasing thermal flux. The FEA models are run using mean values for these quantities appropriate to the operating conditions, and the thermo-electrical model results are insensitive to the variations expected in practice. However, if this level of realism is required and if reliable parametrizations for these dependencies can be obtained, then the inclusion of such variations in the model is possible.

\section{Other model inputs}

The three temperature-dependent elements of the thermo-electrical model---the
radiation-induced digital current increase in the front-end chips, the
efficiency of the FEAST-controlled DC-DC converter, and the sensor leakage current---are described in this section.
Each effect is studied experimentally and fit with functional forms
in order to accurately represent them in the model.
The uncertainty in the experimental data, and in our modelling assumptions,
are estimated here and considered in the evaluation of safety factors,
described in detail in Section~\ref{sec:safety_factors}.

\subsection{DC-DC converter}

The DC-DC converter, which features an air-core inductor and is controlled by the FEAST chip, supplies a low-voltage (1.5~V) current to the ABC130 and HCC front-end
chips on the module.
The efficiency of the FEAST depends on its temperature as well as the output (load) current
load delivered to the front-end chips. To correctly model the FEAST efficiency, experimental
measurements have been performed to characterize the dependence and fitted with a functional form\footnote{
Because of the nearly identical functionality and behaviour expected of the bPOL12V chip
that will replace the FEAST in the final version of the ITk strip system,
the FEAST data and the resulting fit shown here is expected to accurately reflect the bPOL12V case.
}.

For the measurement, the FEAST power board was glued to an aluminum cold plate, cooled
with CO$_2$, and powered with the nominal working input and output voltages (11~V input, 1.5~V output).
The temperature of the FEAST was measured with an NTC (negative temperature coefficient) thermistor and a PTAT (proportional to absolute temperature) sensor residing on the FEAST
for a range of load currents up to the maximum design current of 4A.

\begin{figure}[t!]
\centering
\subfloat[] {\includegraphics[width=0.45\linewidth]{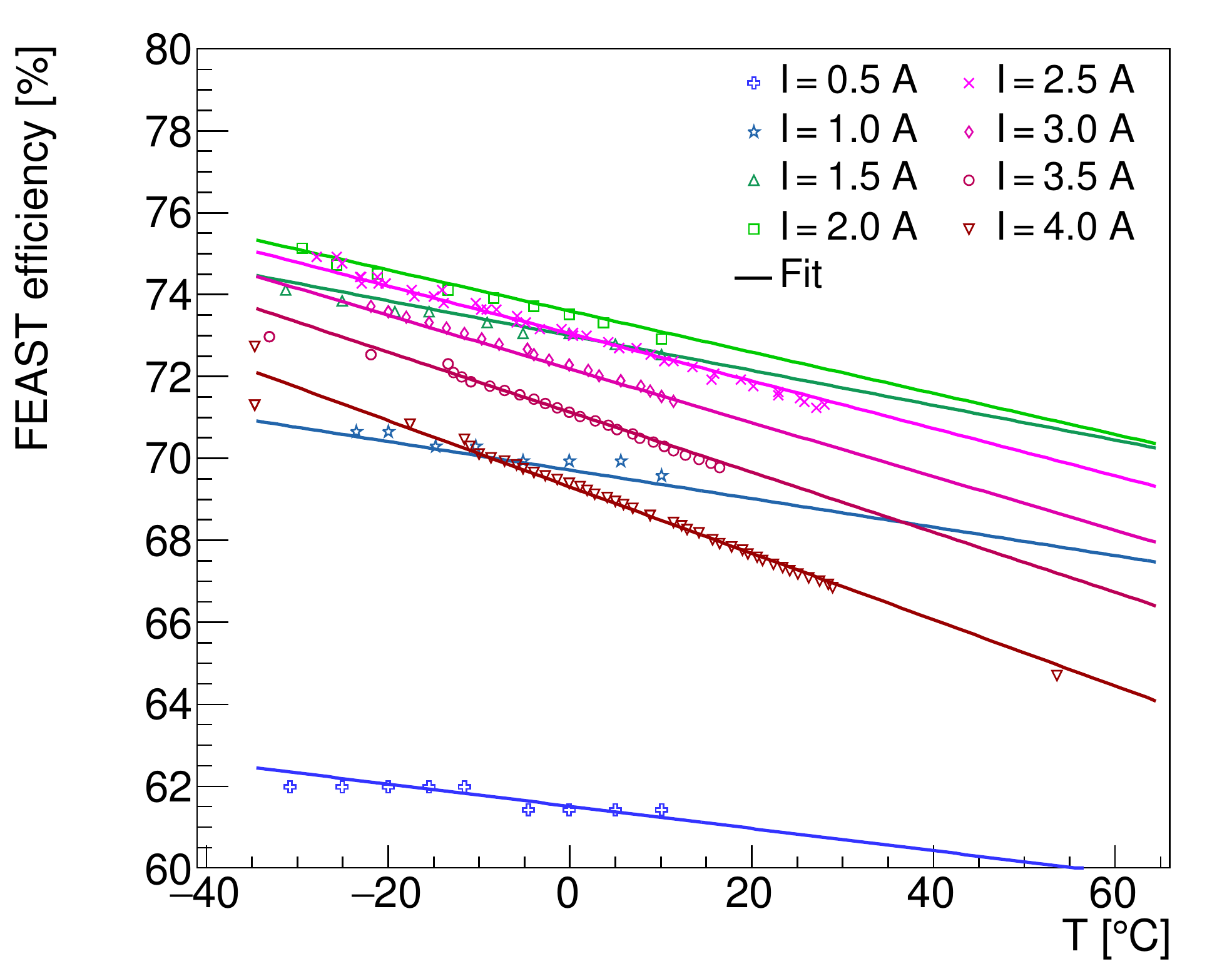}}\quad\quad
\subfloat[] {\includegraphics[width=0.45\linewidth]{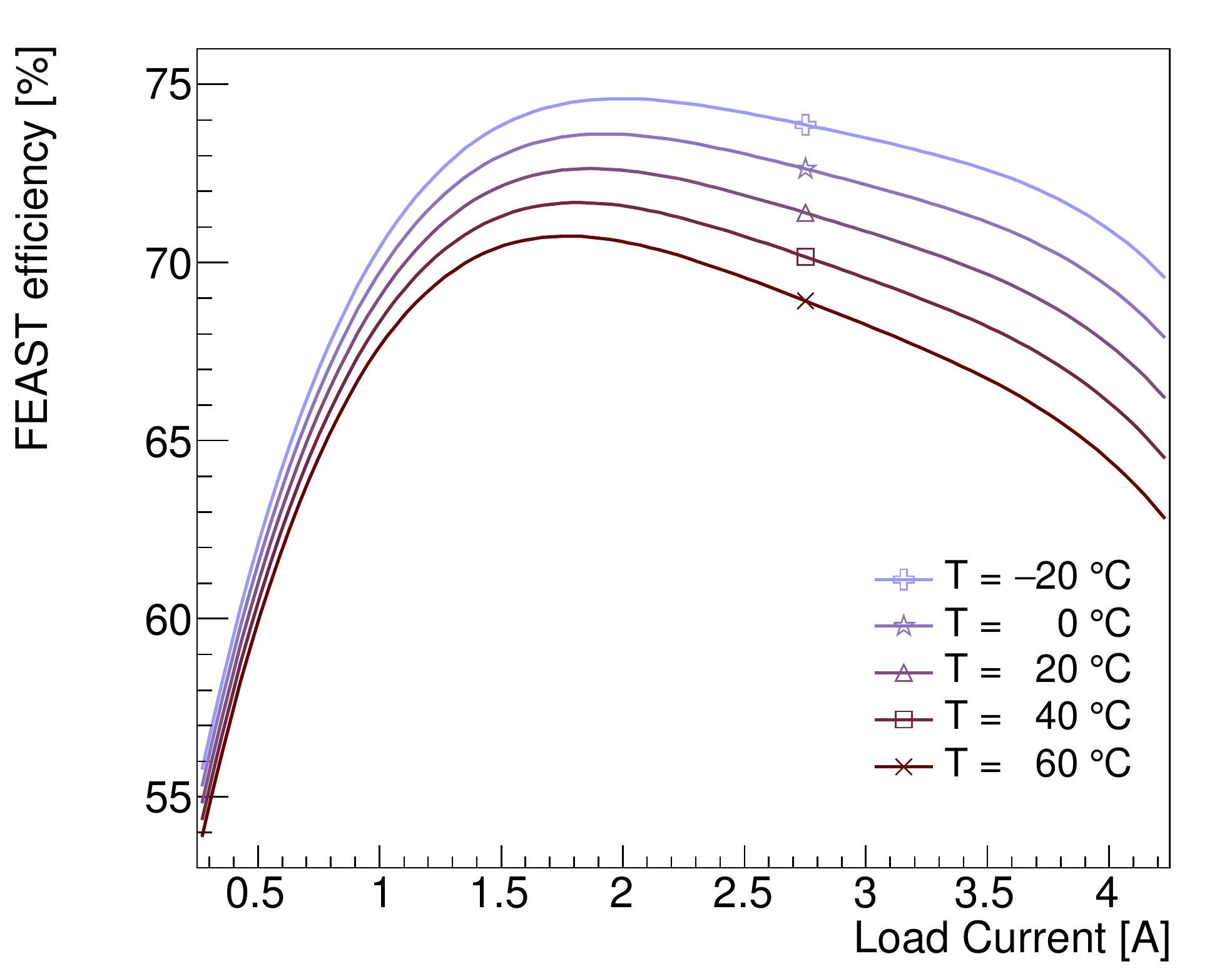}}
\caption{The FEAST efficiency model based on experimental data. (a) The experimental data points
characterizing the FEAST efficiency are plotted as markers and coded by colour and marker style for load current. The data is
compared to the analytic fit, evaluated in curves of equal current. (b) The same analytic fit,
presented as a function of current load for curves of equal temperature.
}
\label{fig:feast_eff}
\end{figure}

The data was then fit with a function with sufficient parameters to ensure reasonable agreement; the
choice of functional form has no physical interpretation. Fig.~\ref{fig:feast_eff} depicts the
FEAST efficiency data and the parametrized fit used in the model. The parametrization fits the data
with an accuracy better than 1\%; this uncertainty in the FEAST efficiency modelling is small
compared to other uncertainty sources, and is therefore neglected in our model.

\subsection{Digital current increase of chips using 130~nm CMOS technology}
\label{sec:tid_explanation}

\begin{figure}[t!]
\centering
\includegraphics[width=0.45\linewidth]{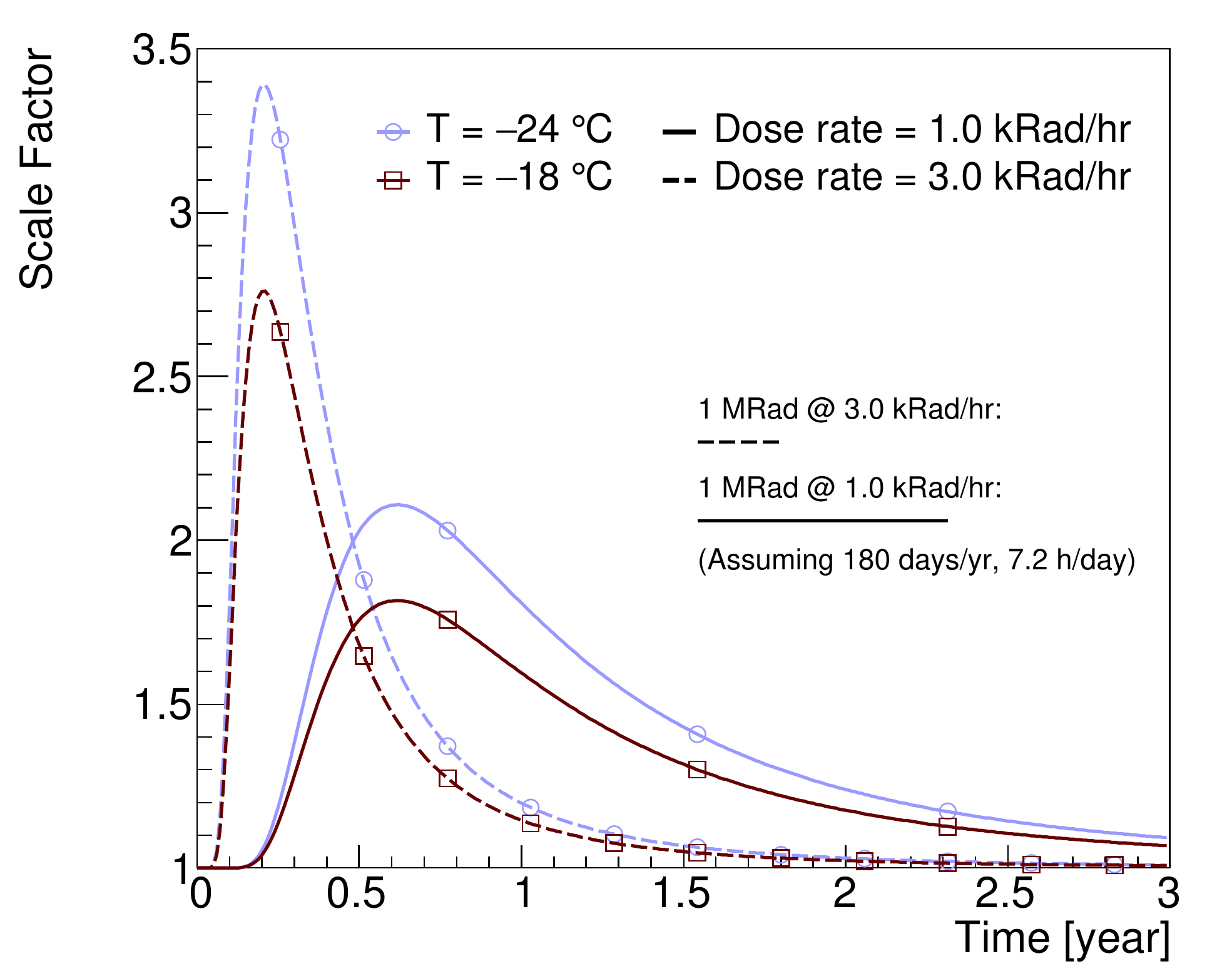}
\caption{
Parameterization of the impact of ionizing radiation
on the magnitude of the front-end chip digital current (the TID bump), presented as a function of time
elapsed during detector operation.
The current is multiplied by a scale factor that is modeled as a function of total ionizing dose,
dose rate, and temperature, using a fit to experimental data.
The figure presents four scenarios using two different chip temperatures and dose rates.
The relationship between time and total ionizing dose is calculated assuming the detector operates
for 180 days per year, 7.2 hours per day.
The time that it takes to reach 1 MRad of total ionizing radiation is indicated on the figure for both dose rates.
}
\label{tid_bump}
\end{figure}

The ABC and HCC chips, designed using IBM 130 nm CMOS 8RF technology, are known to suffer from an
increase in digital current when subjected to a high-radiation environment
\cite{Collaboration:2017mtb}. This phenomenon, known as the `TID bump,' is well-studied
\cite{1589217,FACCIO20081000} and has a characteristic shape whereby the effect reaches a maximum
as a function of the accumulated dose and then gradually diminishes (see Fig.~\ref{tid_bump}).

In an effort to characterize the nature of the TID bump in the ABC and HCC chips empirically,
many irradiation campaigns have been conducted using a variety of radiation sources, testing
the effect at different temperatures and dose rates.
The data collected from these studies was used to develop a model of the TID bump
that estimates the digital current increase given the total ionizing dose, the dose rate,
and the operating temperature of the chip.
The data was fitted using a 5-parameter model which was guided by assumptions on the underlying
damage mechanisms, although the fit function had to be adjusted to achieve a satisfactory match to the data.
The result of this parametrization, which is depicted in
Fig~\ref{tid_bump}, is used as an input to the thermo-electrical model in order to correctly model the
ABC and HCC currents. The TID bump is assumed to fully apply to the HCC digital current, and apply to
69\% of the ABC digital current\footnote{
The primary source of the additional digital current is leakage in a certain type of transistor on the chip,
which is only used in parts of the ABC. We have introduced the fraction factor described in the text
to allow flexibility to scale this leakage current for different chip designs.}.

The TID bump displays certain key features, which are reflected in the parametrization:
first, the effect is larger at colder temperatures and higher dose rates. This means it can be
mitigated by operating the chips at higher temperature (note that the dose rate is determined by the LHC operational conditions).
Second, the figure illustrates how chips receiving different dose rates will reach their maximum
digital current increase at different times. This feature is particularly important when modelling the
total power consumed by the barrel and endcap systems. In both systems, the dose rate varies significantly
depending on the position of the module in the detector. The effect means that the maximum system
power will be smaller than the sum of the maximum power of each module, as each chip reaches
its maximum at a different point in time (see Section~\ref{sec:systemprop}).

The TID bump is an important source of uncertainty in our model. The experimental data exhibit
a relatively large variation in the TID bump effect, in particular
between different batches of the same type of chip delivered by the manufacturer, suggesting an unknown
effect in the fabrication process. To estimate the uncertainty in the TID bump,
the parametrized function is fit again using only the worst-performing data (defined as having the
largest TID bump effect). This `pessimistic' parametrization is used as a safety factor to estimate
the detector performance in worst-case scenarios.

The irradiation of individual chips have typically been performed at constant dose rate and temperature.
However, both of these parameters will vary as a function of time in the scenarios that we attempt to model.
In our current parametrization, we use only the instantaneous value of these two parameters, thus neglecting any possible history of the TID effect for a given chip. We also ignore any short-term effects due to variations in the dose rate on the scale of hours or days. This approach is mandated by the lack of more varied experimental data and the absence of a good theoretical model for this effect. This probably constitutes the largest source of uncertainty in our model.

\subsection{Radiation-dependent leakage current}

The radiation-induced sensor leakage current can be parametrized as a function of the hadron fluence expressed in 1~MeV equivalent neutrons.
We have used linear parametrizations obtained from fits to experimental data taken at 500~V and 700~V at $-15~^\circ$C, and scale them to a given sensor temperature using Eq.~\ref{eq:leakage_current_temp_dependence}.


\section{Running the model}
\label{sec:running}
The thermo-electrical model constructs a profile of the sensor module operation conditions over the
lifetime of the detector in the following manner. First, the total module power (including all components, but excluding the sensor
leakage power) and the sensor temperature assuming no leakage current ($T_0$) are calculated 
using a reasonable set of initial component temperatures.
The initial value for the module power is used to solve for the sensor power and temperature accounting
for leakage current, using the thermal balance equation and the relationship from
Eq.~\ref{eq:leakage_current_temp_dependence}.
Using this calculated sensor leakage current and temperature, the power and temperature of the module
components are updated given the initial (year 0, month 0) startup parameters.

Next, the module conditions of the following month (year 0, month 1) are calculated. Using the component
temperatures calculated from the previous month and the operational parameters (ionizing dose and dose
rates) from the current month, the module total power (excluding sensor leakage) is again calculated, and
subsequently the sensor temperature and leakage current are computed. Following this,
the module component temperatures and power values are derived for this month. This process is repeated in one-month
steps until the final year of operation, or until a real solution for the sensor temperature does not
exist, indicating that thermal runaway conditions have been reached.

In the barrel subsystem, the above procedure is performed four separate times to
represent the radiation conditions of the four barrel layers located at different radii from the beam axis\footnote{The correct module type, short-strip in the inner two layers and long-strip for the outer two layers, is used for each layer.} for both a normal and an EOS-type module. Thus, eight modules are simulated in total for the barrel (4 layers $\times$ normal/EOS), and they are combined in their proper proportion to simulate the entire barrel system.

In the endcap subsystem, the total ionizing dose and dose rates vary significantly depending on the
position of the module; furthermore, the design of each module on a petal differs significantly.
Therefore, all 36 module types (6 rings $\times$ 6 disks) are simulated independently, and combined to
represent the full endcap.

We have implemented this algorithm separately for the barrel and endcaps. In both cases, the calculation for a set of operating conditions over the full lifetime of the LHC, in steps of one month, takes between 5 and 10 minutes on a standard PC. The results represent the equivalent of simulating 1152 barrel modules and 5184 endcap modules, if the same time granularity would be required\footnote{Recall from Section~\ref{sec:impedances} that the thermo-electrical model requires 24 barrel module FEA simulations and 3 endcap petal FEA simulations to extract the thermal impedances, but this procedure is only performed once.}. For comparison, the processing time for a single steady-state FEA simulation of an endcap petal is about 20 minutes (not taking into account the time needed to set the parameters of the simulation). Thus, the thermo-electrical model enables a quick turn-around for systematic studies of the parameter space that would not be possible using FEA alone.

\section{Outputs of the thermo-electrical model}

The thermo-electrical model provides a wide range of predictions for the operation of the strip system. A detailed discussion of all results is beyond the scope of this article; instead, we present here a subset of the results to demonstrate the capabilities and use of the thermo-electrical model for the design of the detector system.

\subsection{Operational scenarios}\label{sec:opscenarios}

\begin{figure}[t!]
\centering
\subfloat[] {\label{fig:lhc_profile}  \includegraphics[width=0.45\linewidth]{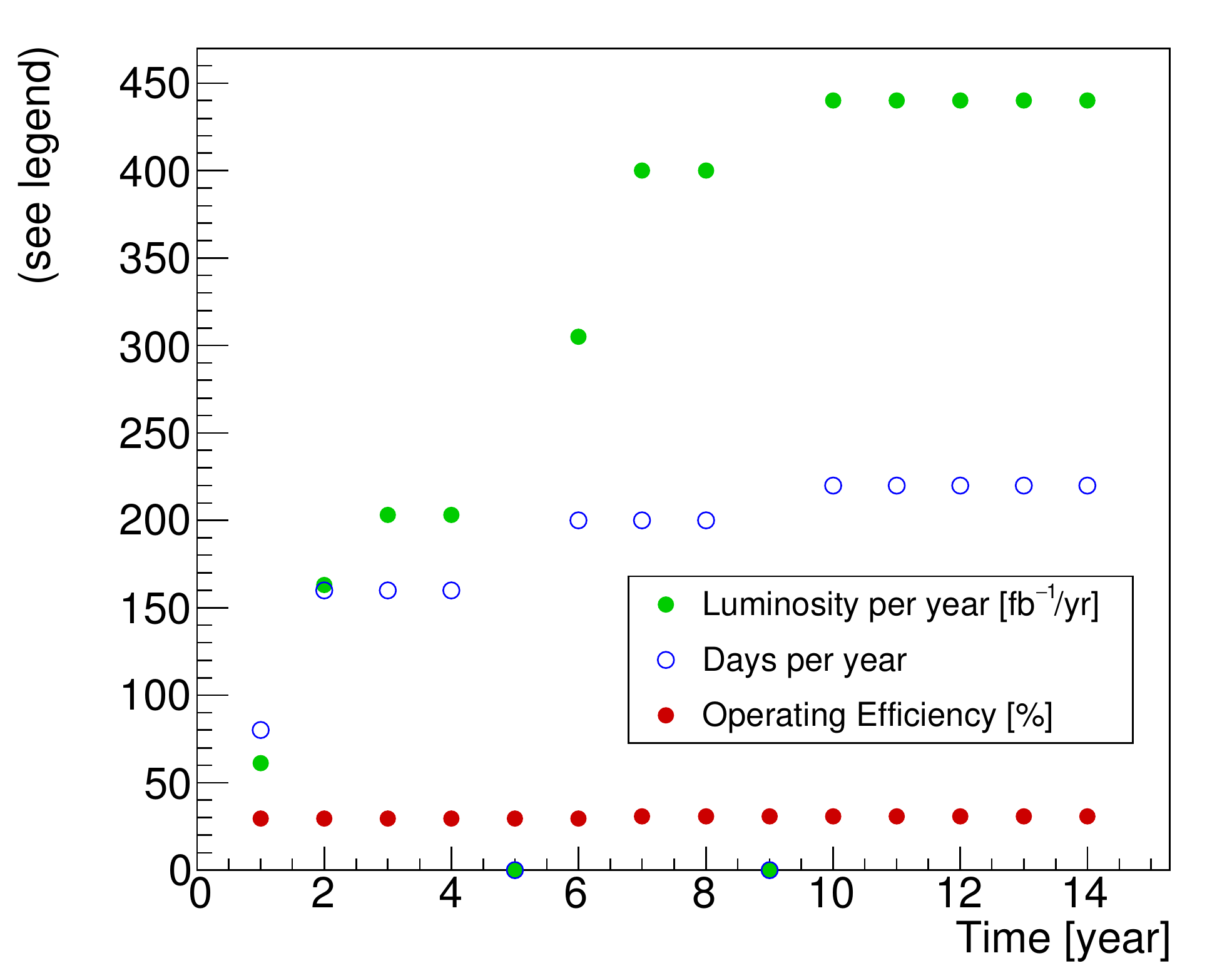}}\quad\quad
\subfloat[] {\label{fig:coolant_ramp} \includegraphics[width=0.45\linewidth]{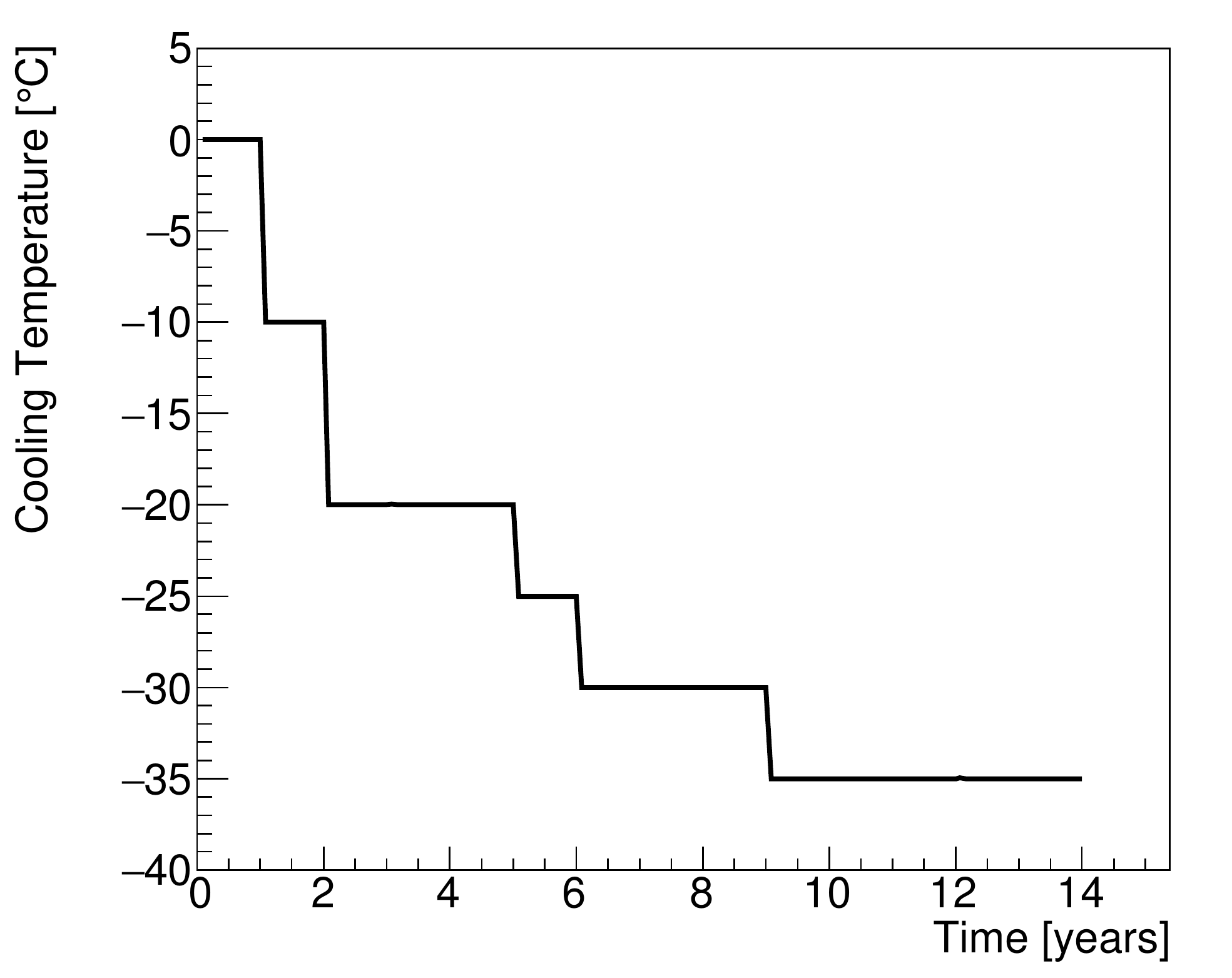}}
\caption{(a) Expected HL-LHC performance and (b) `cooling ramp' scenario for the coolant temperature. Year-long shutdowns of the LHC are anticipated in years 5 and 9.}
\label{fig:opscenarios}
\end{figure}

To study the different aspects of our predictions for the operation of the ITk strip system throughout its lifetime, we performed the calculation of the system parameters over the expected 14 years of operation in monthly steps as outlined in Section~\ref{sec:running}. Time-dependent operational inputs to the calculation were taken from the expected performance of the HL-LHC (Fig.~\ref{fig:lhc_profile}) \cite{ApollinariG.:2017ojx}. For the cooling, which can be adjusted during data taking using detector control systems, we studied flat (constant) coolant temperature profiles ranging from $0~^\circ$C to $-35~^\circ$C, the lowest evaporation temperature achievable with the ITk evaporative CO$_2$ cooling system. We also studied a `ramp' scenario in which the coolant temperature starts at 0~$^\circ$C and is gradually lowered down to $-35~^\circ$C over the course of 10 years (Fig.~\ref{fig:coolant_ramp}).

\subsection{Safety factors}
\label{sec:safety_factors}
\begin{figure}[t!]
\centering
\subfloat[] {\label{fig:safety_factors_a} \includegraphics[width=0.45\linewidth]{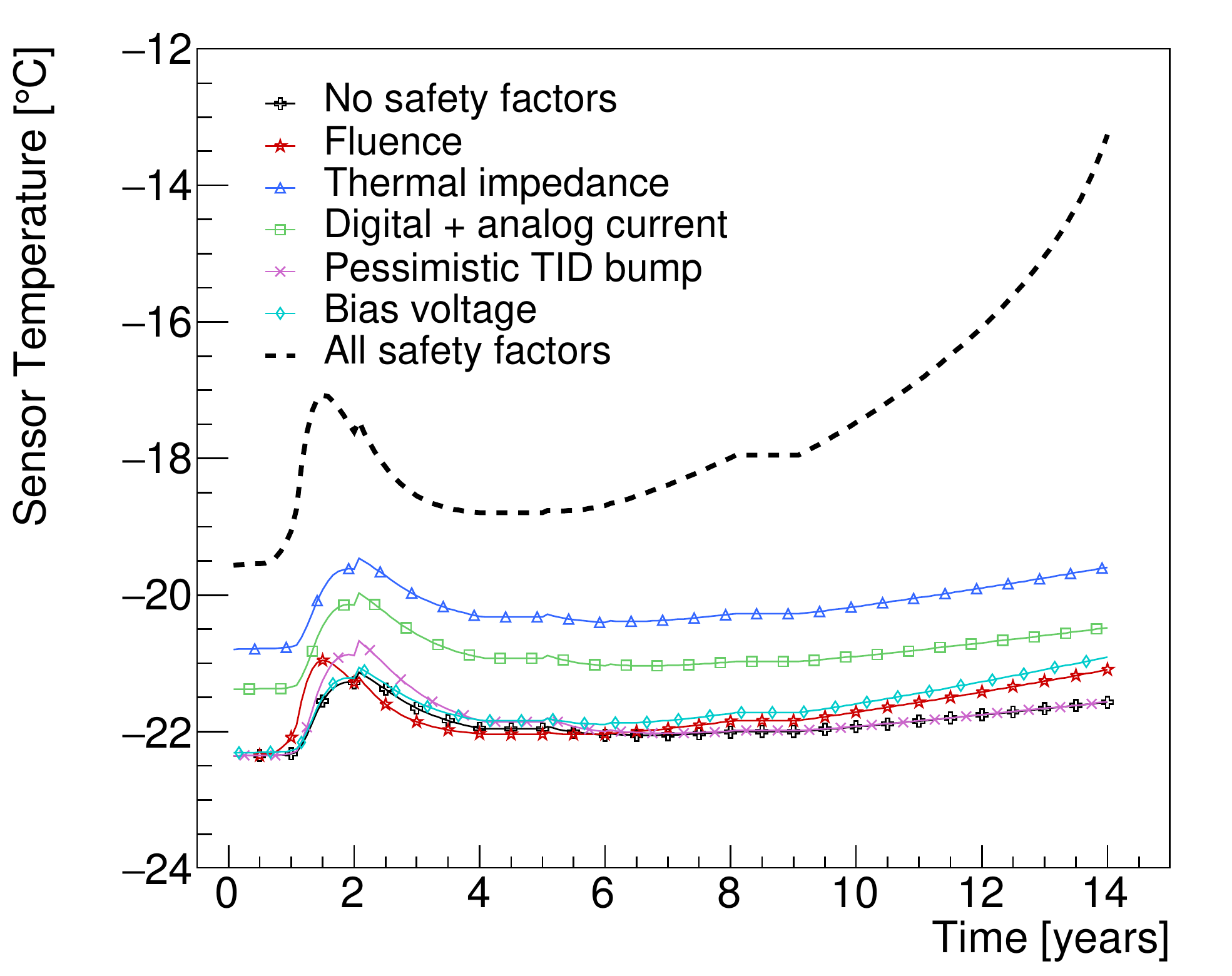}}\quad\quad
\subfloat[] {\label{fig:safety_factors_b} \includegraphics[width=0.45\linewidth]{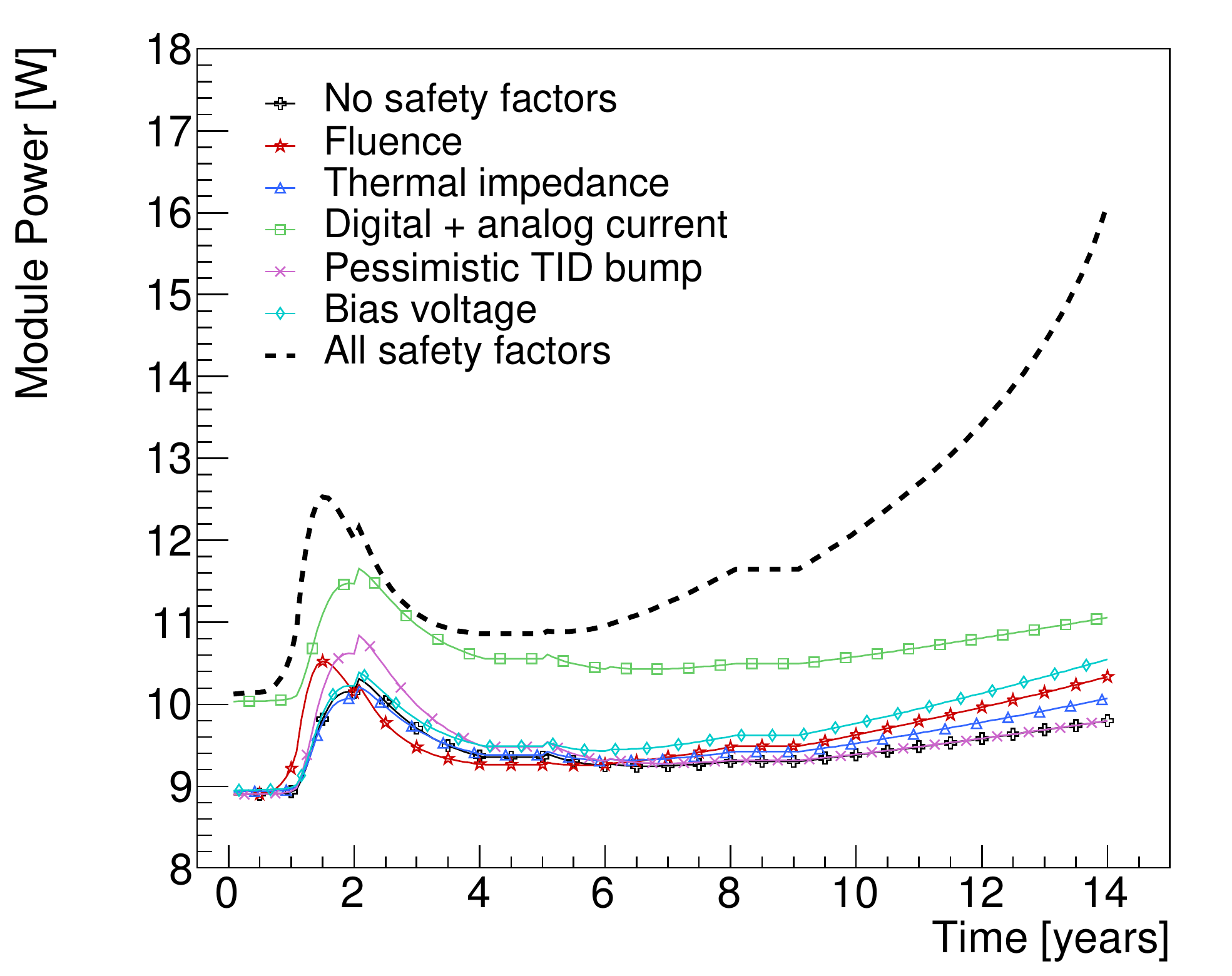}}
\caption{Comparing the impact of different safety factors on (a) the sensor temperature and
(b) the module power for the endcap R3-type module, using a flat cooling scenario ($-30~^\circ$C). The dotted line depicts the effect of all safety
factors applied at once.}
\label{fig:safety_factors}
\end{figure}

To ensure the robustness of the system design against uncertainties in the assumptions used in the model, we also evaluate the model using a set of input parameters with some key inputs degraded. The set of safety factors used is given in Table~\ref{tab:safetyfactors}. Each safety factor has been estimated individually based on experience, the complexity of the system aspect described by the parameter, and from available data or the absence of such data. Note that the model can be evaluated with all the safety factors listed in Table~\ref{tab:safetyfactors} used together, a situation that is unlikely to occur in the real system, to provide a worst-case estimate for the performance of the ITk strip system. The individual effects of the different safety factors are demonstrated in Fig.~\ref{fig:safety_factors}.

\let\arraystretcha\arraystretch 
\renewcommand\arraystretch{1.2} 
\begin{table}[b!]
\caption{Safety factors.}
\label{tab:safetyfactors}
\centering
\adjustbox{max width=\textwidth}{ 
\begin{tabular}{lcl}
\toprule
Safety factor & Value & Reason \\
\midrule
\multirow{2}{*}{Fluence}  & \multirow{2}{*}{50\%} & Accuracy of fluence calculations and uncertainties\\
                          &                       & in material distributions\\
Thermal impedance & 10\% barrel, 20\% endcap & Local support build tolerances, thermal network assumptions\\
Digital current & 20\% & Final chip performance and parametrization of TID effect\\
Analog current & 5\% & Final chip performance\\
Tape electrical impedance & 10\% & Electrical tape manufacturing tolerances\\
Bias voltage & 700~V & Increased bias voltage from nominal 500~V to maintain S/N\\
TID parametrization & Nominal/Pessimistic & Different data sets for fit of TID bump\\
\bottomrule
\end{tabular}
} 
\end{table}
\let\arraystretch\arraystretcha 

It is important to note that combining multiple safety factors can have a compounding effect on the system. As an example, the effect of an increased bias voltage combined with a larger digital current will result in a much higher sensor leakage current at the detector end-of-life than either situation occurring individually. The analytical model presented here allows for scenarios like these to be examined quickly and effectively.

\subsection{Module properties}

\begin{figure}[t!]
\centering
\subfloat[] {\label{fig:moduleflatperformance_a} \includegraphics[width=0.45\linewidth]{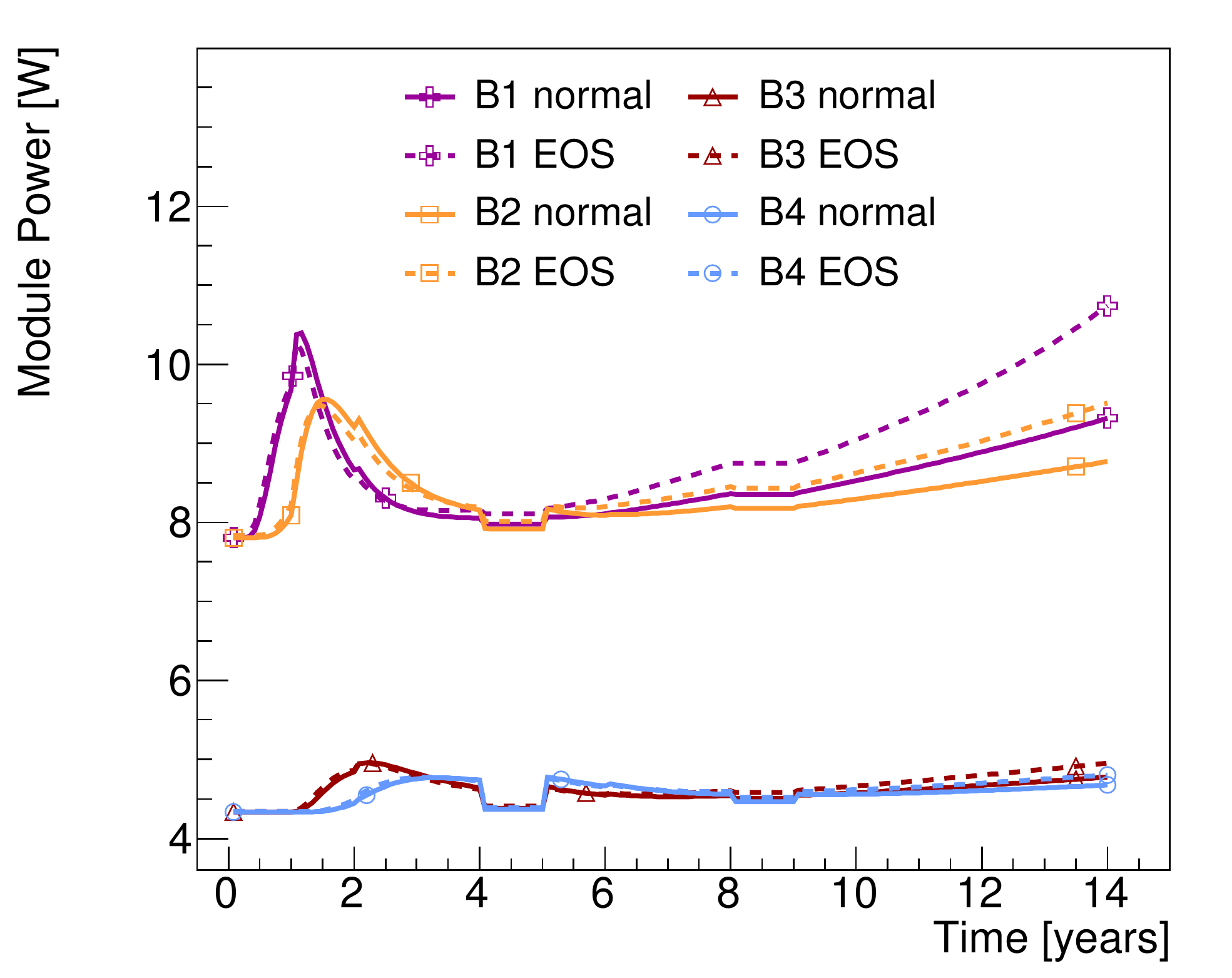}}\quad\quad
\subfloat[] {\label{fig:moduleflatperformance_b} \includegraphics[width=0.45\linewidth]{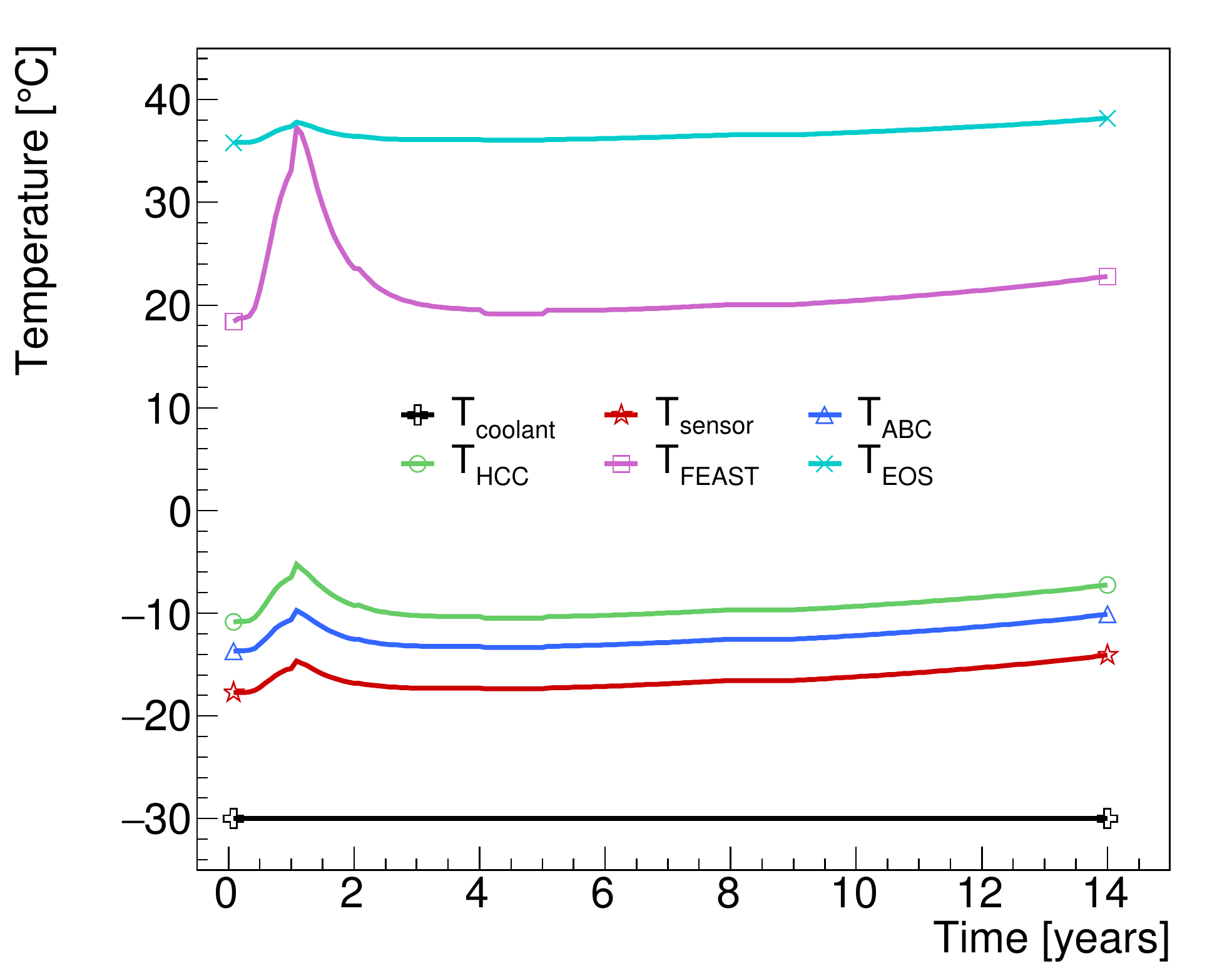}}
\caption{Examples of barrel module performance predictions for a flat cooling scenario ($-30~^\circ$C) including safety factors. (a) Power per module. (b) Temperatures for different nodes of an end-of-stave barrel module in the innermost barrel. The discontinuities in year 5 and 9 are due to anticipated year-long shutdowns of the LHC.}
\label{fig:moduleflatperformance}
\end{figure}

\begin{figure}[t!]
\centering
\subfloat[] {\label{fig:modulerampperformance_a} \includegraphics[width=0.45\linewidth]{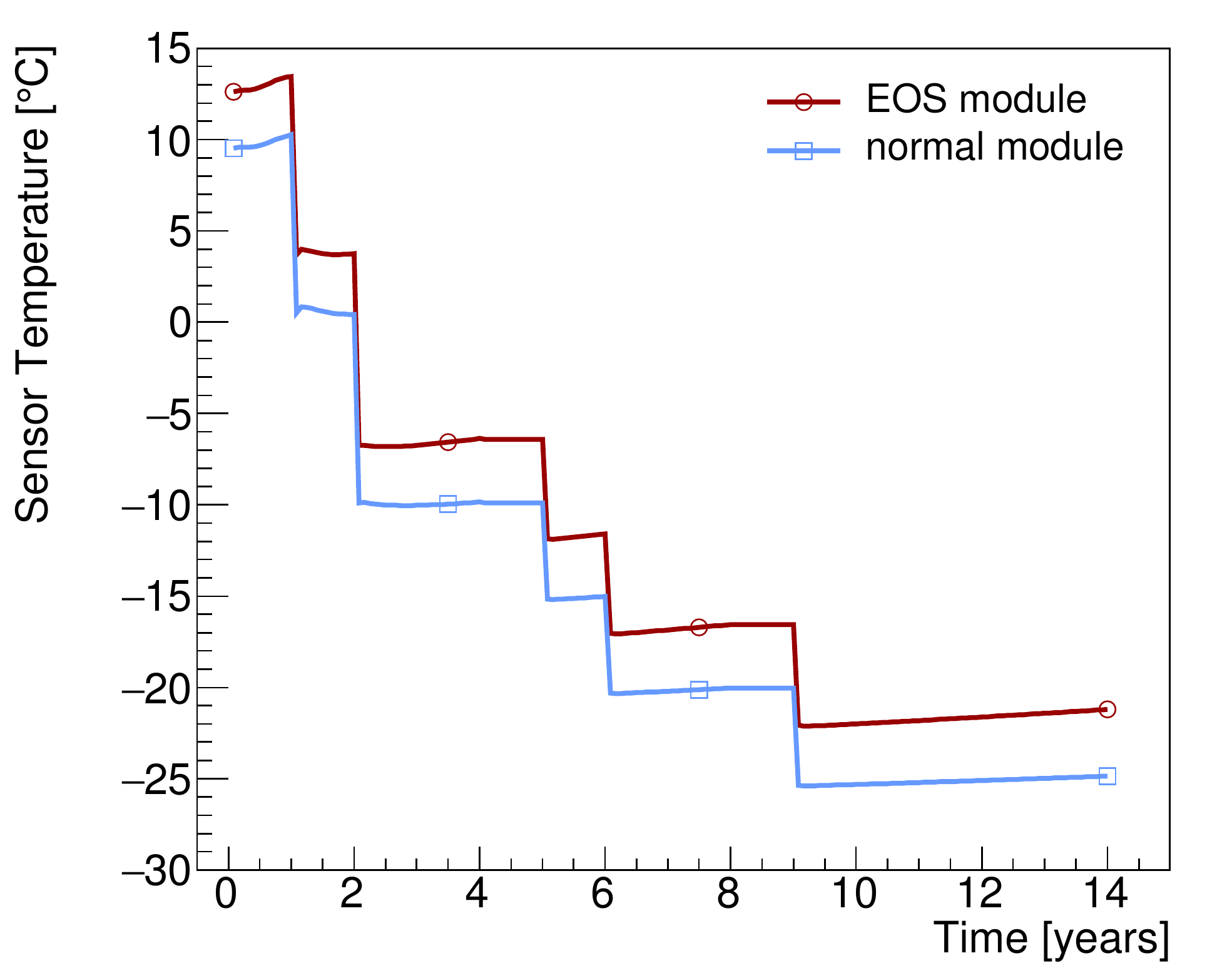}}\quad\quad
\subfloat[] {\label{fig:modulerampperformance_b} \includegraphics[width=0.45\linewidth]{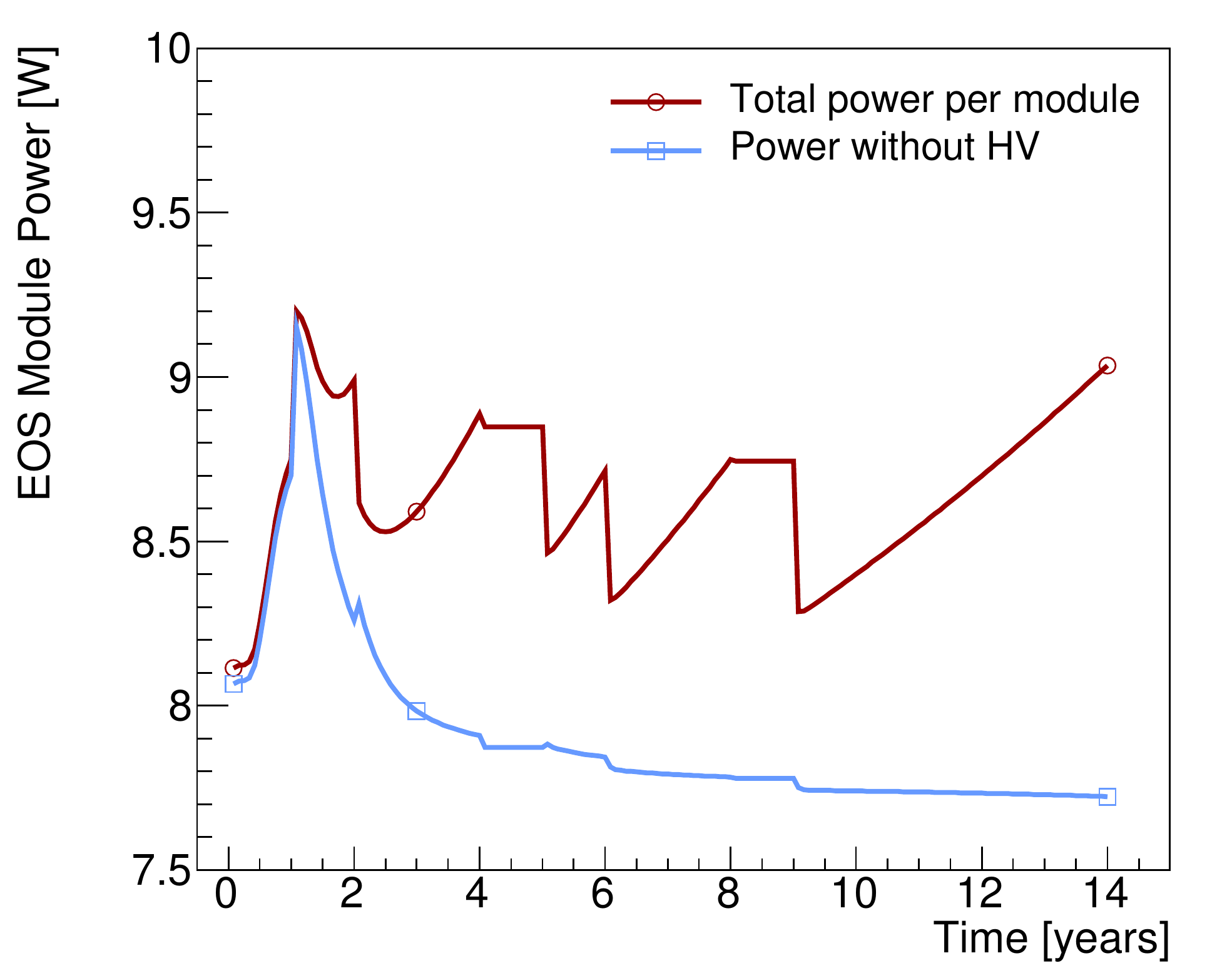}}
\caption{Examples of barrel module performance predictions for the ramp cooling scenario including safety factors. (a) Sensor temperature in the innermost barrel modules. (b) Power in an end-of-stave barrel module in the innermost layer.}
\label{fig:modulerampperformance}
\end{figure}

Several module properties predicted by the thermo-electrical model are shown in Figs.~\ref{fig:moduleflatperformance} and~\ref{fig:modulerampperformance} for the barrel system. The different radiation-dependent effects occur on different timescales. The maximum in the digital chip power due to the TID effect occurs relatively early (in year 1 to 4), although the bump has a long tail, particularly in the outer layers of the barrel. The sensor leakage power, on the other hand, grows towards the end of the lifetime of the ITk. If the leakage current continued to increase in the case of further irradiation, or if the cooling temperature were raised, this growth would ultimately lead to thermal runaway. Due to the radial dependence of the radiation environment, the radiation-induced effects are most pronounced in the innermost barrel layers.

\subsection{System properties}\label{sec:systemprop}
One of the key concerns for the design of the strip system is thermal stability of the system. If the cooling temperature is too high to limit the leakage power from the radiation-damaged sensors to a level where the heat can still be removed, the system is unstable (it goes into `thermal runaway').
To find the cooling temperature $T_\text{C}$ at which this condition is reached, we make repeated simulations of the ITk strip system using the thermo-electrical model, with each simulation representing the full 14-year operation of the ITk at a fixed $T_\text{C}$. Between simulations, $T_\text{C}$ is increased in steps of 5~$^\circ$C until the model finds thermal runaway. In the numeric evaluation of the thermo-electrical model this manifests itself in the absence of a solution to the system of equations. In the endcap strip system, this occurs at a cooling temperature of $-15~^\circ$C under nominal conditions (i.e. with no safety factors applied); in this scenario, thermal runaway would be reached in the 12$^\text{th}$ year of operation. With all safety factors applied, thermal runaway would occur at a cooling temperature of $-25~^\circ$C (in year 10).
In the barrel system, where the radiation environment is slightly less intense, the conditions for thermal runaway occur at the same cooling temperatures, but a few years later than in the endcaps: in the final year of operation and a cooling temperature of $-15~^\circ$C under nominal conditions, and at $-25~^\circ$C (in year 13) with safety factors applied.
As the design cooling temperature of the ITk cooling system is $-35~^\circ$C, we have confidence that the ITk strip system has a sufficient margin for thermal stability.

\begin{figure}[t!]
\centering
\subfloat[] {\label{fig:systemperformance_a} \includegraphics[width=0.45\linewidth]{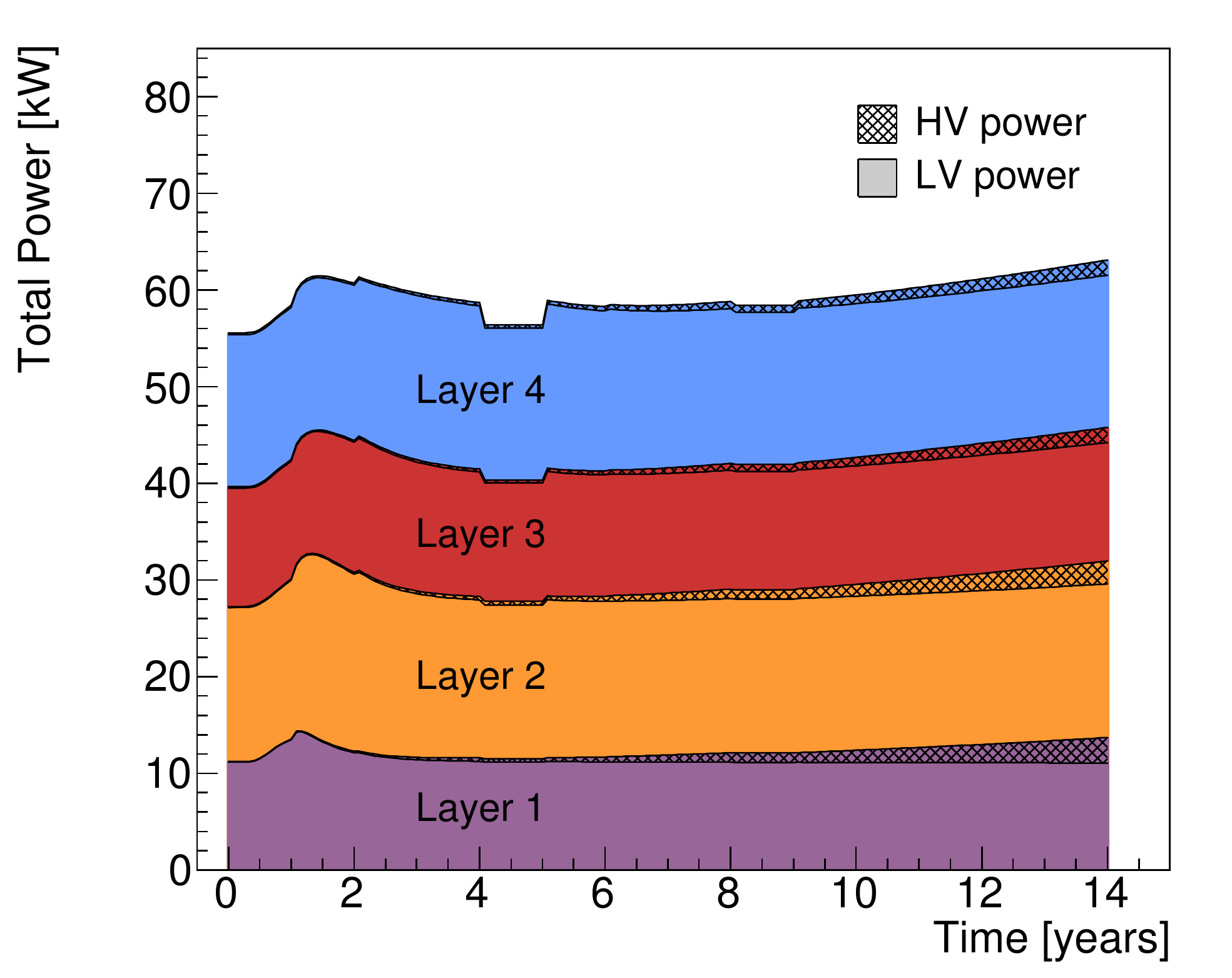}}\quad\quad
\subfloat[] {\label{fig:systemperformance_b} \includegraphics[width=0.45\linewidth]{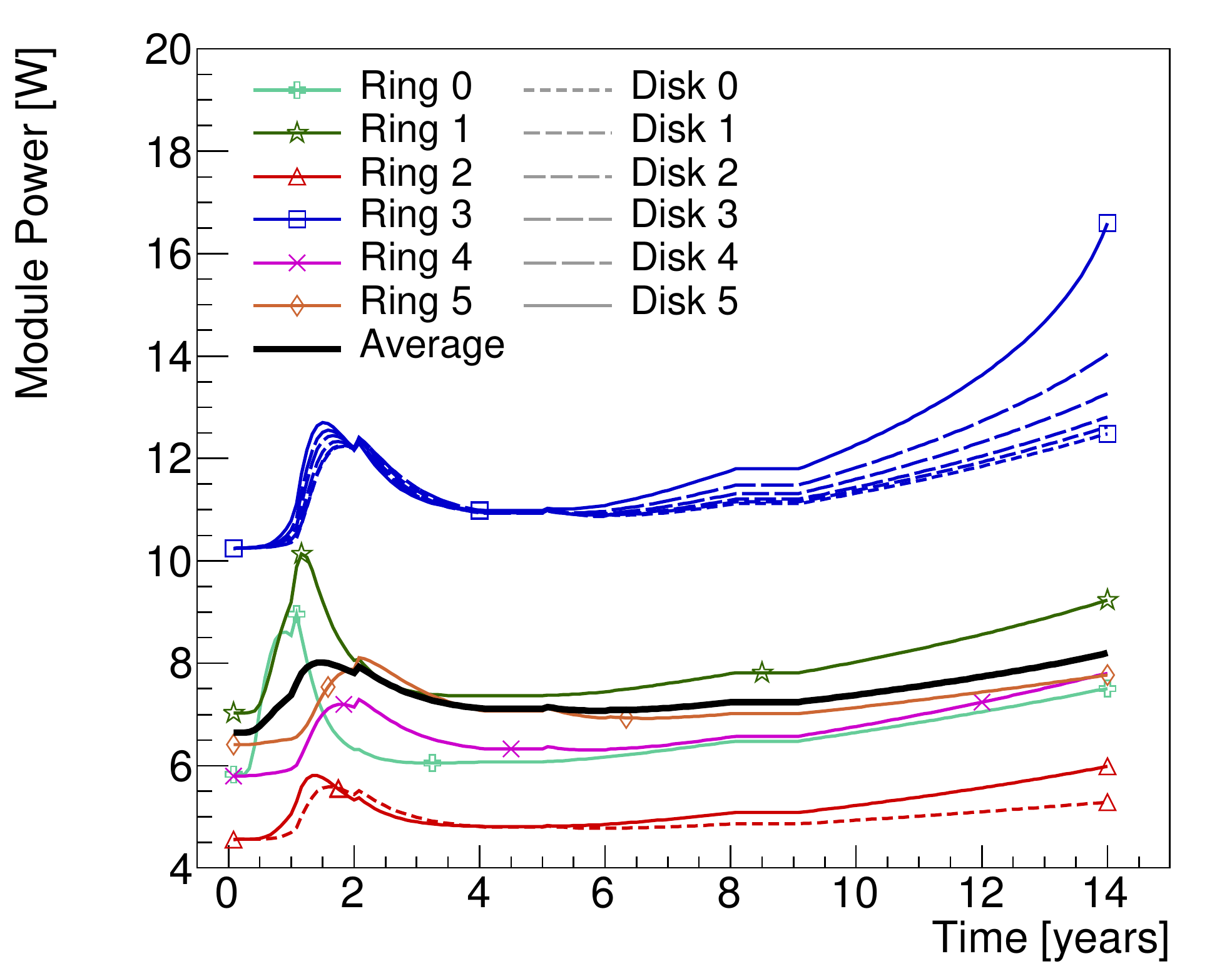}}
\caption{Examples of system performance predictions. (a) Barrel total power requirements. The plot shows the stacked power requirements for the four barrel layers (purple: layer 1, orange: layer 2, red: layer 3, blue: layer 4). Full colour indicates power from the front-end electronics, and hatched parts are contributions from HV power for the four barrels. The discontinuities in year 5 and 9 are due to anticipated year-long shutdowns of the LHC. (b) The power requirements for 12 of the 36 simulated endcap modules, labelled according to their ring type and disk position. (Some modules are omitted to improve the clarity of the figure.) The solid black line indicates the average module power. Both predictions use a scenario with flat $-30~^\circ$C cooling and including all safety factors.}
\label{fig:systemperformance}
\end{figure}

Beyond the issue of stability, the thermo-electrical model delivers predictions for the development of current and power requirements for the overall system. Some of the predictions are shown in Fig.~\ref{fig:systemperformance}. Again, the different timescales of the various radiation-induced effects are visible; ignoring this time dependence could lead to over-specification of some system aspects.
Taking Fig.~\ref{fig:systemperformance_b} as an example, the average module power (indicated by the thick black line) is 6.6~W in the beginning of operation and reaches 8.0~W at the TID bump, in the second year. If we naively summed the maxima of the TID bumps, neglecting time dependence, we would arrive at 8.6~W, overestimating the power at the TID bump by 7.5\% and overestimating the effect of the TID bump by 43\%.
The difference, multiplied in the endcap by 4608 modules, amounts to nearly 3~kW of power and impacts the specifications of e.g. the cooling system.

The predictions from this model are now used throughout the strip project to consistently size the power supply and cooling systems. Including safety factors in the predictions gives us some confidence that the designs are robust; by using commonly agreed safety factors, we ensure a consistent use of safety factors throughout the project and prevent safety factor creep.

\begin{figure}[t!]
\centering
\subfloat[] {\label{fig:rampoptimization_a} \includegraphics[width=0.45\linewidth]{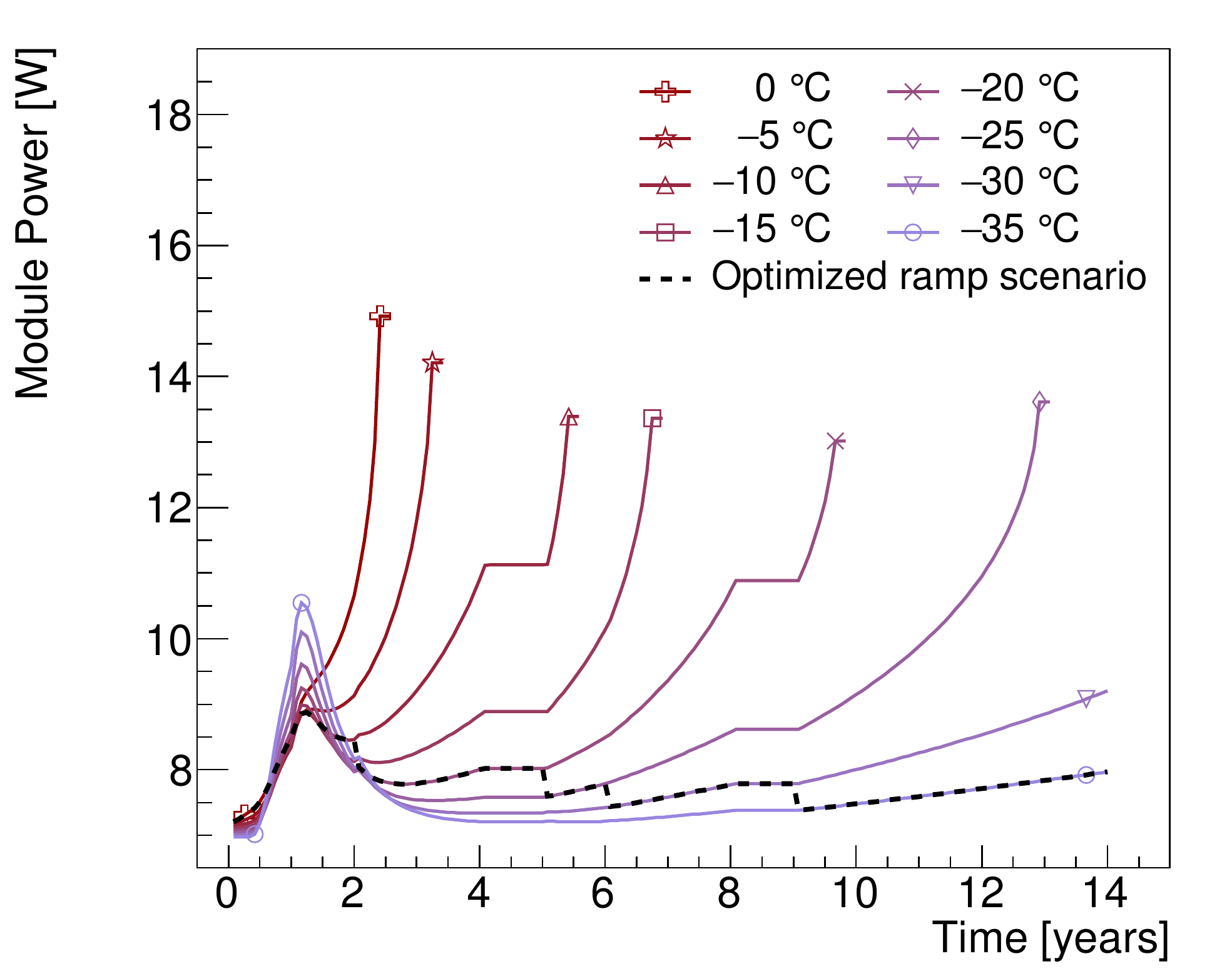}}\quad\quad
\subfloat[] {\label{fig:rampoptimization_b} \includegraphics[width=0.45\linewidth]{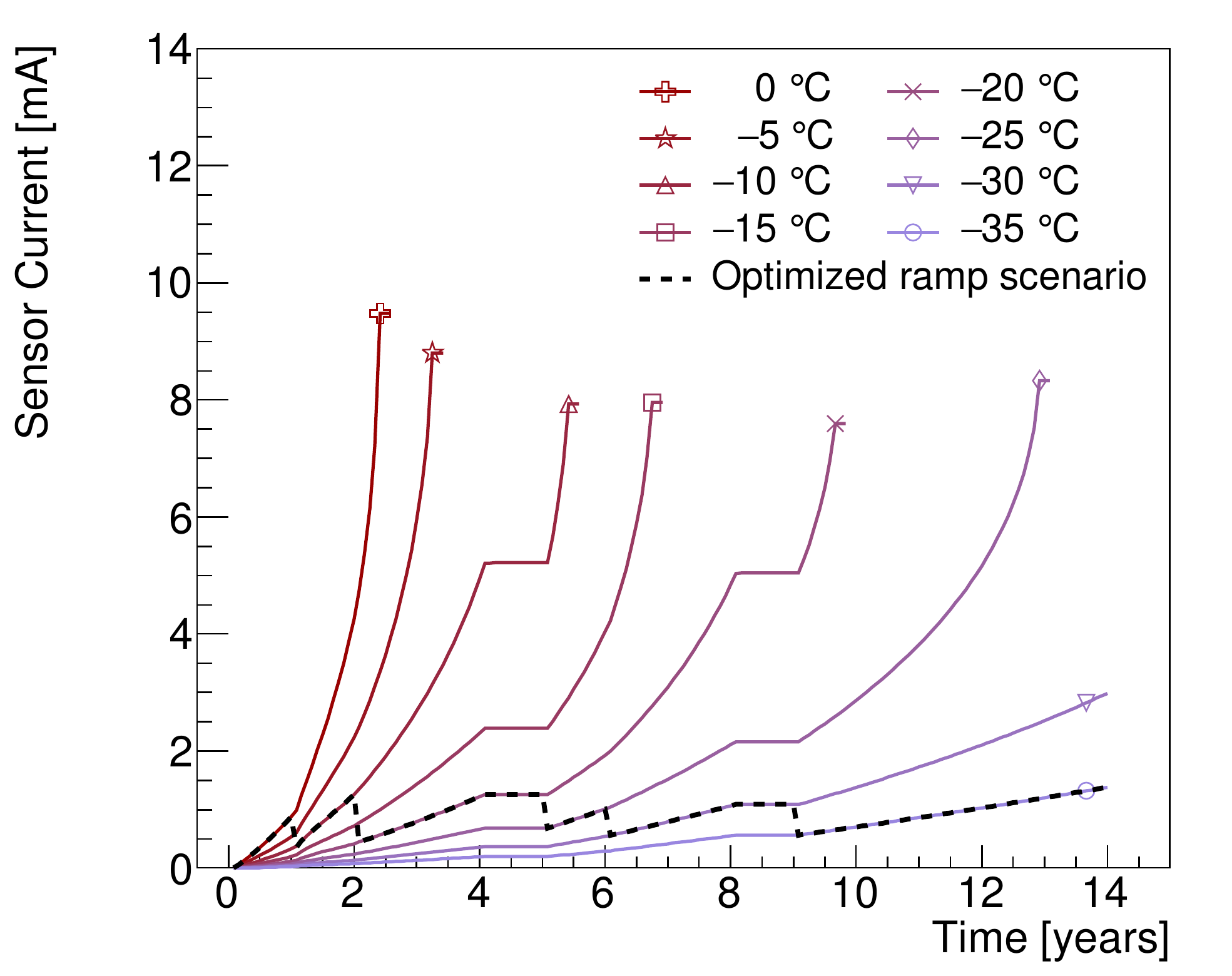}}
\caption{(a) Total power and (b) sensor leakage current of the endcap R1-type module for eight different flat cooling profiles, ranging from 0~$^\circ$C to $-35~^\circ$C, as well as the cooling ramp scenario specified in Fig.~\ref{fig:coolant_ramp} (dashed curve). The curves that are discontinued before year 14 correspond to scenarios that have reached thermal runaway. The cooling ramp scenario has been selected to minimize the module power while keeping the sensor leakage current stable throughout the lifetime of the ITk. All safety factors are applied in these plots.}
\label{fig:rampoptimization}
\end{figure}

Because of the different timescales for the peak power due to the TID effect and the radiation-induced sensor leakage, there is room to optimize the cooling temperature profile to minimize the total power in the strip system while avoiding thermal runaway. The thermo-electrical model is a powerful tool to plan such an optimized cooling profile. In fact, the cooling `ramp' scenario introduced in Section~\ref{sec:opscenarios} is the result of such an optimization.
In this scenario, depicted in Fig.~\ref{fig:rampoptimization}, the cooling temperature begins at a relatively high value ($0~^\circ$C) to minimize the impact of the TID bump in the first two years of operation, thus avoiding a peak in the module power (see Fig.~\ref{fig:rampoptimization_a}). In subsequent years, $T_C$ is steadily decreased to maintain a sensor current at or below about 1~mA, as illustrated in Fig.~\ref{fig:rampoptimization_b}, in the interest of both minimizing the module power and avoiding thermal runaway.

\section{Model performance verification}
The accuracy of the predictions of the thermo-electrical model is affected by two major factors: the quality of the input parameters, and the error introduced by reducing the complex 3D geometry into a linear thermal impedance network. The former has been discussed throughout this paper where the different inputs have been presented. For the latter, we have studied the agreement of predictions from the network model with the more accurate results obtained from FEA for selected states of the system.

To verify the level of this agreement, we have calculated the sensor temperature curve for a barrel EOS-type module up to thermal runaway, both in the full FEA and in the network model. For this exercise, we do not vary any of the input parameters in the model other than the sensor leakage power with its temperature dependence. The resistor values in the network model are the same as used throughout for our model, obtained as described in Section~\ref{sec:impedances}. For the power from the various electronics components, the FEAST efficiency and the TID scale factor we have used representative nominal values.

\begin{figure}[t!]
\centering
\subfloat[] {\label{fig:verification_a} \includegraphics[width=0.11\linewidth,valign=c]{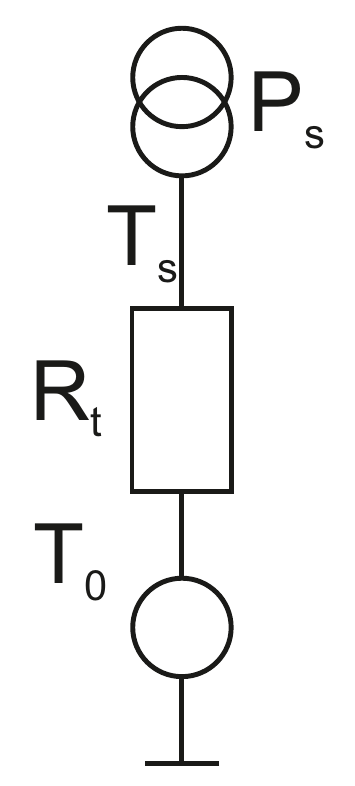}}\quad
\subfloat[] {\label{fig:verification_b} \includegraphics[width=0.32\linewidth,valign=c]{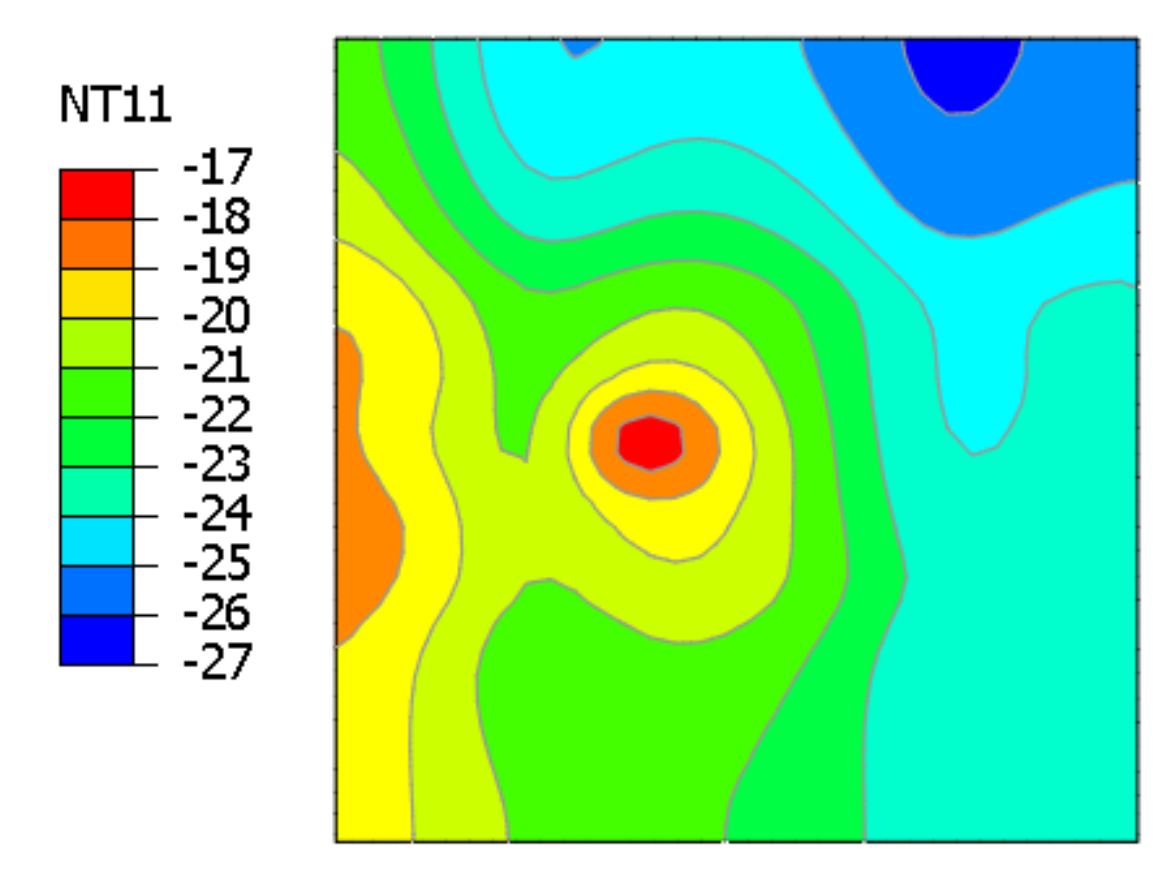}}\quad\quad
\subfloat[] {\label{fig:verification_c} \includegraphics[width=0.45\linewidth,valign=c]{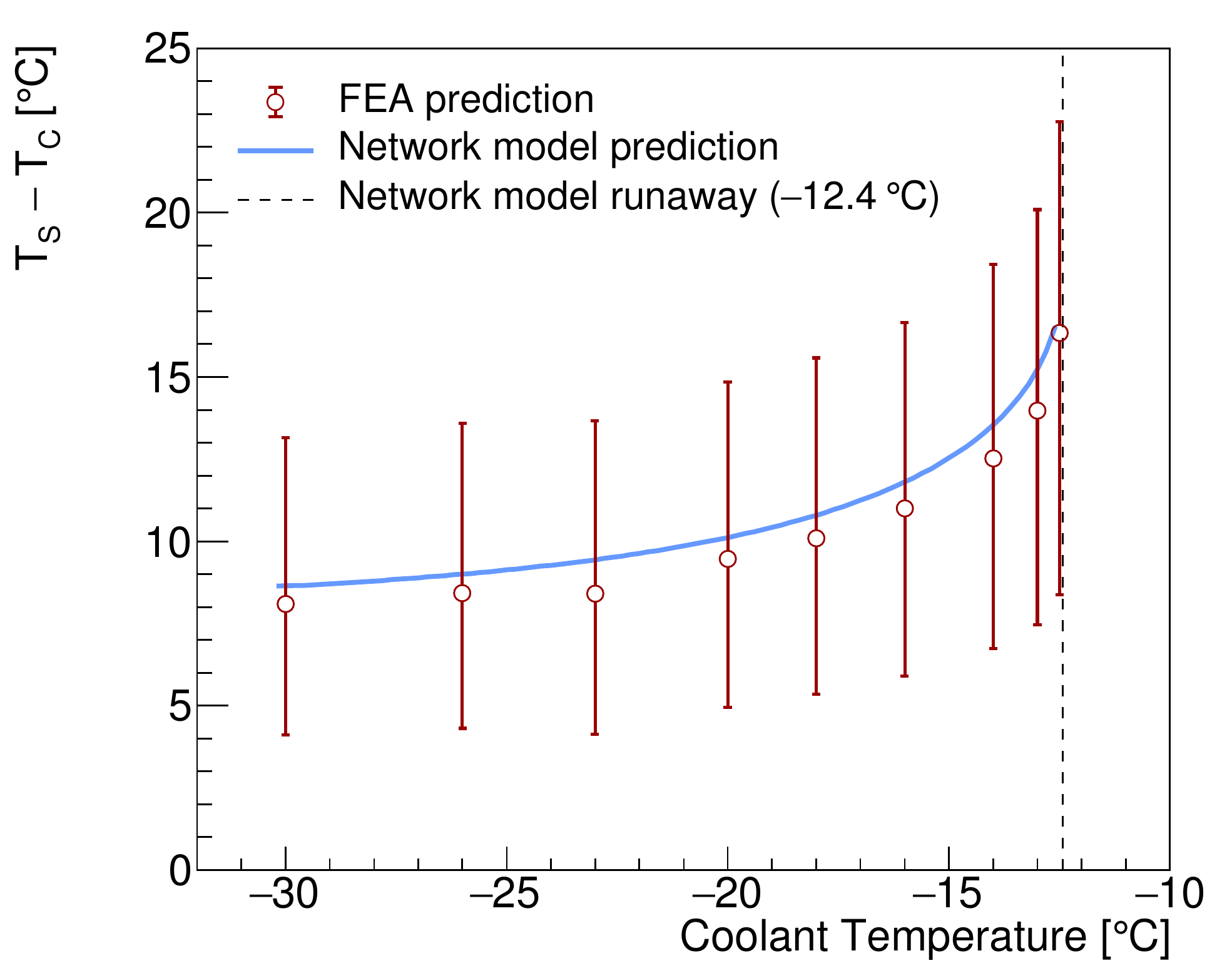}}
\caption{(a) Th\'{e}venin equivalent of the thermal network. (b) Result of sensor surface temperature calculations using FEA, assuming zero sensor power. The EOS card is to the left of the module, and the cooling pipes run from top to bottom about a quarter of the module width from each edge. The figure uses a rainbow colour gradient, with blue indicating the lowest temperatures and red the highest temperatures. (c) Difference of average sensor and coolant temperature, comparing FEA (red points) and the network model prediction (blue curve). The bars on the FEA data indicate minimum and maximum sensor temperature. The dotted vertical line indicates the critical temperature derived analytically using the network model ($-12.4~^\circ$C).}
\label{fig:verification}
\end{figure}

Because the variable model inputs are kept constant for this study, we can reduce the complex thermal network to its Th\'{e}venin equivalent, which is identical to the network studied in Ref.~\cite{Beck:2010zzd}, and use the analytical expressions given there. The reduced network is described by the base temperature $T_0$, defined as the sum of the coolant temperature and the temperature rise due to the front-end electronics alone, and the total thermal impedance $R_t$ from the sensor to the coolant. Using the nominal resistances and representative power numbers from the module, $T_0=-21.9~^\circ$C and $R_t=1.132$~K/W in the network model, compared to $-22.4~^\circ$C and 1.147~K/W obtained directly from the FEA. The comparison of the predicted sensor temperatures for both cases is shown in Fig.~\ref{fig:verification}. Despite a large temperature variation of about 10~$^\circ$C across the sensor, the network model runaway prediction agrees well with the FEA\footnote{The critical temperature here is $-12.4~^\circ$C, which is higher than the numbers given in Section~\ref{sec:systemprop}, because the study here ignores temperature effects such as the FEAST efficiency, which can only be modelled in the network model.}. This gives us confidence that the use of a thermal network model is not likely to significantly degrade the predictions beyond the uncertainties introduced by other inputs to the model. 

\section{Conclusions}
We have developed a model of the ATLAS ITk strip system that is based on the interplay between a thermal and an electrical network model. The set of equations in the model can be numerically solved using standard data analysis software in a short time, allowing for a quick turn-around for systematic studies of the system performance. The complexity of these networks is given by the number of interconnected components between the networks, many of which have a non-linear dependence on the temperature or electrical power. This approach can be easily adopted for any other silicon detector system.

In the case of the ATLAS strip system, several temperature-dependent heat sources had to be modeled. In addition to the sensor leakage current, these are the  radiation-induced increase of the digital front-end power (`TID bump') and the efficiency of the DC-DC conversion system. The outputs of the model give us confidence that the ITk strip system will be thermally stable until the end of LHC Phase-II operation, even with the inclusion of safety factors on key inputs. Furthermore, the model provides information for benchmark system parameters like cooling, supply power and currents in power cables, which is used in the specification of these systems. The use of the model outputs throughout the strip project ensures consistent specifications, including a common strategy on safety factors. Using the thermo-electrical model, we can also propose an optimized cooling temperature `ramp' scenario, which stabilizes leakage power throughout the lifetime of the experiment while minimizing the TID bump.

We have verified the performance of the thermal network model compared to a full FEA treatment, and we are confident that the level of disagreement is smaller than the uncertainty introduced by the model inputs. Among the inputs, the most likely source of uncertainty stems from the limitations in our understanding of the parametrization of the TID effect.

\section{Acknowledgements}
The evaluation of the thermo-electrical model depends critically on the input parameters to the model. To capture the whole of the system, these need to distill all that is known of the system, and we are therefore indebted to the whole of the ITk strip community. In particular, we would like to thank Tony Affolder, Kyle Cormier, Ian Dawson, Sergio Diez Cornell, Laura Gonella, Ashley Greenall, Alex Grillo, Paul Keener, Steve McMahon, Paul Miyagawa, Craig Sawyer, Francis Ward and Tony Weidberg for all their inputs to this work.

The research was supported and financed in part by the UK's Science and Technology Facilities Council and the Helmholtz Association (HGF) in Germany.

\bibliographystyle{elsarticle/elsarticle-num}
\bibliography{paper}

\begin{thebibliography}{10}
\expandafter\ifx\csname url\endcsname\relax
  \def\url#1{\texttt{#1}}\fi
\expandafter\ifx\csname urlprefix\endcsname\relax\def\urlprefix{URL }\fi
\expandafter\ifx\csname href\endcsname\relax
  \def\href#1#2{#2} \def\path#1{#1}\fi

\bibitem{Chilingarov_2013}
A.~Chilingarov,
  \href{https://doi.org/10.1088\%2F1748-0221\%2F8\%2F10\%2Fp10003}{{Temperature
  dependence of the current generated in Si bulk}}, Journal of Instrumentation
  8~(10) (2013) P10003--P10003.
\newblock \href {https://doi.org/10.1088/1748-0221/8/10/p10003}
  {\path{doi:10.1088/1748-0221/8/10/p10003}}.
\newline\urlprefix\url{https://doi.org/10.1088\%2F1748-0221\%2F8\%2F10\%2Fp10003}

\bibitem{Collaboration:2017mtb}
A.~Collaboration, \href{https://cds.cern.ch/record/2257755}{{Technical Design
  Report for the ATLAS Inner Tracker Strip Detector}}, Tech. Rep.
  CERN-LHCC-2017-005, ATLAS-TDR-025, CERN, Geneva (Apr 2017).
\newline\urlprefix\url{https://cds.cern.ch/record/2257755}

\bibitem{ATL-INDET-PUB-2017-001}
\href{https://cds.cern.ch/record/2291800}{{Radiation induced effects in the
  ATLAS Insertable B-Layer readout chip}}, Tech. Rep. ATL-INDET-PUB-2017-001,
  CERN, Geneva (Nov 2017).
\newline\urlprefix\url{https://cds.cern.ch/record/2291800}

\bibitem{1748-0221-6-11-C11035}
A.~Affolder, B.~Allongue, G.~Blanchot, F.~Faccio, C.~Fuentes, A.~Greenall,
  S.~Michelis, \href{http://stacks.iop.org/1748-0221/6/i=11/a=C11035}{{DC-DC
  converters with reduced mass for trackers at the HL-LHC}}, Journal of
  Instrumentation 6~(11) (2011) C11035.
\newline\urlprefix\url{http://stacks.iop.org/1748-0221/6/i=11/a=C11035}

\bibitem{dcdc-info}
\href{https://project-dcdc.web.cern.ch/project-dcdc/}{{Development of DCDC
  converters @ CERN}} [cited 2020-02-17].
\newline\urlprefix\url{https://project-dcdc.web.cern.ch/project-dcdc/}

\bibitem{Beck:2010zzd}
G.~Beck, G.~Viehhauser, {Analytic model of thermal runaway in silicon
  detectors}, Nucl. Instrum. Meth. A618 (2010) 131--138.
\newblock \href {https://doi.org/10.1016/j.nima.2010.02.264}
  {\path{doi:10.1016/j.nima.2010.02.264}}.

\bibitem{background}
\href{https://twiki.cern.ch/twiki/bin/view/AtlasPublic/RadiationSimulationPublicResults#FLUKA_Simulations}{{ATLAS
  Experiment - Radiation Simulation Public Results}} [cited 2020-02-17].
\newline\urlprefix\url{https://twiki.cern.ch/twiki/bin/view/AtlasPublic/RadiationSimulationPublicResults#FLUKA_Simulations}

\bibitem{abc130}
N.~Lehmann,
  \href{https://documents.epfl.ch/users/n/nl/nlehmann/www/SelfSeededTrigger_MasterThesis/SelfSeededTrigger_NiklausLehmann_Thesis.pdf}{{Tracking
  with self-seeded Trigger for High Luminosity LHC}}, Master's thesis, Section
  of Electrical and Electronical Engineering, {\'E}cole Polytechnique
  F{\'e}d{\'e}rale de Lausanne, Lausanne Switzerland (2014).
\newline\urlprefix\url{https://documents.epfl.ch/users/n/nl/nlehmann/www/SelfSeededTrigger_MasterThesis/SelfSeededTrigger_NiklausLehmann_Thesis.pdf}

\bibitem{abaqus}
M.~Smith, {ABAQUS/Standard User's Manual, Version 6.9}, Simulia, 2009.

\bibitem{ansys}
{ANSYS, Inc.}, \href{http://www.ansys.com/}{{ANSYS Academic Research
  Mechanical, Release 18.2}}.
\newline\urlprefix\url{http://www.ansys.com/}

\bibitem{1589217}
F.~Faccio, G.~Cervelli, {Radiation-induced edge effects in deep submicron CMOS
  transistors}, IEEE Transactions on Nuclear Science 52~(6) (2005) 2413--2420.
\newblock \href {https://doi.org/10.1109/TNS.2005.860698}
  {\path{doi:10.1109/TNS.2005.860698}}.

\bibitem{FACCIO20081000}
F.~Faccio, H.~J. Barnaby, X.~J. Chen, D.~M. Fleetwood, L.~Gonella, M.~McLain,
  R.~D. Schrimpf,
  \href{http://www.sciencedirect.com/science/article/pii/S0026271408000826}{{Total
  ionizing dose effects in shallow trench isolation oxides}}, Microelectronics
  Reliability 48~(7) (2008) 1000 -- 1007, 2007 Reliability of Compound
  Semiconductors (ROCS) Workshop.
\newblock \href
  {https://doi.org/https://doi.org/10.1016/j.microrel.2008.04.004}
  {\path{doi:https://doi.org/10.1016/j.microrel.2008.04.004}}.
\newline\urlprefix\url{http://www.sciencedirect.com/science/article/pii/S0026271408000826}

\bibitem{ApollinariG.:2017ojx}
G.~Apollinari, I.~B{\'e}jar~Alonso, O.~Br{\"u}ning, P.~Fessia, M.~Lamont,
  L.~Rossi, L.~Tavian, {High-Luminosity Large Hadron Collider (HL-LHC)}, CERN
  Yellow Rep. Monogr. 4 (2017) 1--516.
\newblock \href {https://doi.org/10.23731/CYRM-2017-004}
  {\path{doi:10.23731/CYRM-2017-004}}.

\end{thebibliography}

\end{document}